\documentclass[fleqn,usenatbib]{mnras}

\usepackage{newtxtext,newtxmath}

\usepackage[T1]{fontenc}
\usepackage{ae,aecompl}

\usepackage{graphicx}	
\usepackage{amsmath}	
\usepackage{bm}
\usepackage{adjustbox}

\usepackage{hyperref}

\def\apjl{ApjL}

\newcommand{\bh}{\bm{\hat{b}}}

\newcommand{\pc}{p_c}

\newcommand{\vrm}{{\rm v}}
\newcommand{\bvrm}{\bm{{\rm v}}}

\newcommand{\vstdp}{{\rm v}_{{\rm st}, \Delta p}}
\newcommand{\bvstdp}{\bm{{\rm v}_{{\rm st}, \Delta p}}}

\newcommand{\nub}{\nu_{\rm B}}

\title[CR buoyancy instability]{A new buoyancy instability in galaxy clusters due to streaming cosmic rays}

\author[Kempski, Quataert \& Squire]{
Philipp Kempski,$^{1}$\thanks{E-mail: philipp.kempski@berkeley.edu}
Eliot Quataert,$^{2}$
Jonathan Squire$^{3}$
\\
$^{1}$Department of Astronomy and Theoretical Astrophysics Center, University of California, Berkeley, CA 94720, USA \\
$^{2}$Department of Astrophysical Sciences, Princeton University, Princeton, NJ 08544, USA \\
$^{3}$Department of Physics, University of Otago, 730 Cumberland St, North Dunedin, Dunedin 9016, New Zealand}
\date{Accepted XXX. Received YYY; in original form ZZZ}
\pubyear{2020}

\begin{document}

\label{firstpage}
\pagerange{\pageref{firstpage}--\pageref{lastpage}}
\maketitle

\begin{abstract}
Active Galactic Nuclei (AGN) are believed to provide the energy that prevents runaway cooling of gas in the cores of galaxy clusters. However, how this energy is transported and thermalized throughout the Intracluster Medium (ICM) remains unclear. In recent work we showed that streaming cosmic rays (CRs) destabilise sound waves in dilute ICM plasmas. Here we show that CR streaming in the presence of gravity also destabilises a pressure-balanced wave. We term this new instability the CR buoyancy instability (CRBI). In stark contrast to standard results without CRs, the pressure-balanced mode is highly compressible at short wavelengths due to CR streaming. Maximal growth rates are of order $(p_c / p_g) \beta^{1/2} \omega_{\rm ff}$, where $p_c/p_g$ is the ratio of CR pressure to thermal gas pressure, $\beta$ is the ratio of thermal to magnetic pressure and $\omega_{\rm ff}$ is the free-fall frequency. The CRBI operates alongside buoyancy instabilities driven by background heat fluxes, i.e. the heat-flux-driven buoyancy instability (HBI) and the magneto-thermal instability (MTI). When the thermal mean free path $l_{\rm mfp}$ is $\ll$ the gas scale height $H$, the HBI/MTI set the growth rate on large scales, while the CRBI sets the growth rate on small scales. Conversely, when $l_{\rm mfp} \sim H$ and $(p_c/p_g) \beta^{1/2} \gtrsim 1$, CRBI growth rates exceed HBI/MTI growth rates even on large scales. Our results suggest that CR-driven instabilities may be partially responsible for the sound waves/weak shocks and turbulence observed in galaxy clusters. CR-driven instabilities generated near radio bubbles may also play an important role redistributing AGN energy throughout clusters.

\end{abstract}

\begin{keywords}
cosmic rays -- galaxies: clusters: intracluster medium -- galaxies: evolution -- instabilities -- plasmas
\end{keywords}

\section{Introduction}
The cores of galaxy clusters are filled with virialized, hot gas, with typical temperatures exceeding $10^7$ K. The X-ray luminosities of most cluster cores imply cooling times that are significantly shorter than the ages of these systems. Without a source of heating, this hot gas is expected to cool, sink to the center and form stars at a high rate. However, observations find significantly smaller star formation rates and cold gas masses than are predicted by the ``cooling flow" model (e.g., \citealt{pf06}). This suggests that there is a source of heating present that keeps the gas in cluster cores in approximate thermal balance.   

Central Active Galactic Nuclei (AGN) and the interaction of their jets with the Intracluster Medium (ICM) are believed to play an important role in providing the energy that  prevents runaway cooling of ICM gas. In particular, observations suggest that energy is carried away from the central AGN by jet-inflated bubbles of relativistic plasma that buoyantly rise into the ICM. There is a strong correlation between the power needed to inflate the bubbles and the radiative losses of the hot gas (\citealt{churazov2000}; \citealt{birzan2004}; \citealt{rafferty2006}; \citealt{nulsen2009}; \citealt{hlavacek-larrondo2012}; see \citealt{werner2019} for a recent review). 

How this energy is subsequently transported and thermalized throughout cluster cores remains an open question. It is possible that the buoyantly rising radio bubbles stir turbulence by exciting internal gravity waves (IGWs; e.g., \citealt{zhuravleva_2016}, \citealt{zhang_2018}), launch sound waves and/or weak shocks (e.g., \citealt{fabian03}, \citealt{fabian06}, \citealt{2009_sternberg_soker}), and/or inject cosmic rays (CRs) into the ICM (e.g., \citealt{guo08}, \citealt{jp_1}, \citealt{jp_2}). These processes can plausibly occur to some extent simultaneously, but it is unclear which (if any) one is the dominant channel for ICM heating.

Relativistic CRs from both star formation and AGN may play an important role in the evolution of gas in clusters by driving outflows and heating diffuse gas (e.g., \citealt{bmv91}; \citealt{loew91}; \citealt{everett08}; \citealt{socrates08}; \citealt{guo08} \citealt{zweibel_micro}; \citealt{ruszkowski17}; \citealt{zweibel17}; \citealt{jp_1}; \citealt{jp_2}; \citealt{ehlert18}; \citealt{farber18}; \citealt{kq2020}; \citealt{qtj_2021_streaming}; \citealt{qtj_2021_diff}). CRs couple to the thermal gas by scattering from small-scale magnetic fluctuations. In self-confinement theory, cosmic rays are scattered by Alfv\'en waves propagating down the CR pressure gradient, which they themselves excite through the streaming instability (\citealt{kp69}). Pitch-angle scattering by the excited Alfv\'en waves isotropises the cosmic rays in the frame of the waves. In the absence of damping of the self-excited waves, this results in CR streaming relative to the thermal gas at the local Alfv\'en speed. If damping is present, CR propagation deviates from pure Alfv\'enic streaming. The form of the transport correction is, however, quite peculiar, as it corresponds to neither streaming nor diffusion (\citealt{skilling71}, \citealt{wiener2013}; \citealt{kq2022}).

In recent work, we showed that streaming cosmic rays drive a rapidly growing acoustic instability in dilute ICM plasmas, the Cosmic Ray Acoustic Braginskii (CRAB) instability (\citealt{kqs2020}). This suggests that the different channels for transferring and thermalizing energy in the ICM (waves, CRs, turbulence...), usually considered separately in theoretical models, may in fact be closely related. Here, we demonstrate that in the presence of gravity, streaming cosmic rays also destabilise a pressure-balanced wave, more specifically the CR entropy mode modified by gravity. We term this instability the CR buoyancy instability (CRBI) because CRs and buoyancy (gravity) are critical for setting its properties. The growth rates of the CRBI are of order the natural buoyancy frequency (the local free-fall frequency) for plausible ICM parameters. Our work demonstrates the potential physical richness of CR feedback in dilute plasmas: both the CRBI considered in this work and the CRAB instability in \cite{kqs2020} are driven by CR streaming at the Alfv\'en speed, which in a weakly collisional plasma depends on the pressure anisotropy of the thermal gas (the pressure is anisotropic because of the large thermal-particle mean free path in the ICM). This dependence introduces a new form of coupling between the CRs and the thermal gas, which is very unstable.

The remainder of this work is organised as follows. We present our model of cosmic rays coupled to a low-collisionality plasma  in Section \ref{sec:model}. In Section \ref{sec:prelims} we provide a physical overview of the instability. We derive a dispersion relation and asymptotic growth rates in Section \ref{sec:grav}. In Section \ref{sec:icm} we present numerical solutions to the linearised system of equations for an isothermal background.  We discuss the relationship to other buoyancy instabilities and thermal instability, and the importance of CR diffusion in Section \ref{sec:discussion}. We summarise our results in Section \ref{sec:conclusions}.

\section{Model}{\label{sec:model}}
\subsection{Equations}{\label{sec:eqns}}
We consider a  low-collisionality plasma coupled to streaming cosmic rays. We use the weakly-collisional Braginskii MHD model to describe the thermal gas (\citealt{br65}). The equations for the gas and cosmic rays are, 
\begin{equation} \label{eq:cont}
\frac{\partial \rho}{\partial t} +  \bm{\nabla \cdot} ( \rho \bvrm ) = 0,
\end{equation}
\begin{equation} \label{eq:mom}
\rho \frac{d \bvrm}{d t} = - \bm{\nabla} \Big( p_\perp + p_c + \frac{B^2}{ 8 \pi} \Big) + \frac{\bm{B \cdot \nabla B}}{4 \pi} + \bm{\nabla \cdot } \big( \bm{\hat{b} \hat{b}} \Delta p \big) + \rho \bm{g},
\end{equation}
\begin{equation} \label{eq:induction}
\frac{\partial \bm{B}}{\partial t} = \bm{\nabla \times} (\bm{\bvrm \times B}),
\end{equation}
\begin{equation}\label{eq:s}
\rho T \frac{ds}{dt} = -\bm{\vstdp \cdot \nabla}p_c - \bm{\Pi : \nabla \vrm}   - \bm{\nabla \cdot Q} - \rho^2 \Lambda(T),
\end{equation}
\begin{equation} \label{eq:pc}
\frac{dp_c}{dt} = -\frac{4}{3}p_c \bm{\nabla \cdot} ( \bvrm + \bvstdp) - \bm{\vstdp\cdot \nabla}p_c + \bm{\nabla \cdot} \big(\kappa \bm{\hat{b}\hat{b} \cdot \nabla} p_c \big),
\end{equation}
where  $\bvrm$ is the gas velocity, $\rho$ is the gas density, $\bm{B}$ is the magnetic field (with unit vector $\bh$), $s=k_{\rm B} \ln(p_g/\rho^{\gamma}) / (\gamma -1)m_{\rm H}$ is the gas entropy per unit mass, $\Lambda(T)$ is the temperature-dependent cooling function and $\pc$ is the CR pressure. $d / dt \equiv \partial / \partial t + \bm{{\rm v} \cdot \nabla}$ denotes a total (Lagrangian) time derivative. $\Delta p = p_\perp - p_\parallel$ is the gas pressure anisotropy, where $p_\perp$ and $p_\parallel$ denote the pressures in the directions perpendicular and parallel to the magnetic field, respectively. $p_\perp$ and $p_\parallel$ are related to $p_g$ in eq. \eqref{eq:s} by
\begin{equation} \label{eq:ptot}
    p_\perp = p_g + \frac{1}{3} \Delta p.
\end{equation}
The pressure anisotropy in Braginskii MHD  (with viscosity $\nub \approx p_g / \rho \nu_{\rm ii}$, where $\nu_{\rm ii}$ is the ion-ion collision rate; \citealt{br65}) is given by
\begin{equation} \label{eq:Deltap}
    \Delta p = p_\perp - p_\parallel = 3 \rho \nub \big( \bm{\hat{b} \hat{b} : \nabla \vrm} - \frac{1}{3} \bm{\nabla \cdot \vrm} \big). 
\end{equation}
The viscous stress tensor in the gas-entropy equation is
\begin{equation}
    \bm{\Pi} = - \Delta p \Big(\bm{\hat{b}\hat{b} - \frac{\mathcal{I}}{3}} \Big).
\end{equation}
In the absence of background $\Delta p$, the perturbed $\bm{\Pi : \nabla \vrm} $ in the gas-entropy equation is second-order and does not contribute in our linear analysis. $\bm{Q}$ in equation \eqref{eq:s} is the anisotropic thermal heat flux, 
\begin{equation}
    \bm{Q} = - \kappa_{\rm B} \bm{\hat{b}\hat{b}\cdot \nabla}T,
\end{equation}
where $\kappa_{\rm B}$ is the thermal conductivity.

$\vstdp$ in equation \eqref{eq:pc} is the CR streaming speed.  We assume that cosmic rays stream down their pressure gradient at the Alfv\'en speed, which in low-collisionality plasmas depends on the thermal-gas pressure anisotropy,\footnote{We note that the original expression for the Alfv\'en speed in \cite{kqs2020} was incorrect, as the 1/2 exponent in the $\Delta p$ term in \eqref{eq:va_mod} was by accident omitted. This was corrected in \cite{kqs_erratum}. However, the conclusions of \cite{kqs2020} are not affected by this change.}
\begin{equation} \label{eq:va_mod}
    \bvstdp = \chi \frac{\bm{B}}{\sqrt{4 \pi \rho}} \Big(1 + \frac{4\pi \Delta p}{B^2} \Big)^{1/2},
\end{equation}
where $\chi \equiv  -\bm{\hat{b} \cdot \nabla} p_c / |\bm{\hat{b} \cdot \nabla} p_c| = \pm 1$ ensures that the cosmic rays stream along $\bm{B}$ down their pressure gradient and makes the CR heating term $-\bm{{\rm v}_{\rm st} \cdot \nabla}p_c$ in the gas energy equation (\ref{eq:s}) positive definite. We also include CR diffusion along the magnetic field in eq. \eqref{eq:pc}.

\subsection{Dimensionless parameters and characteristic frequencies} \label{sec:params}
We define the ratio of CR pressure to gas pressure,
\begin{equation} \label{eq:eta}
 \eta \equiv \frac{p_c}{p_g}   ,
\end{equation}
and the ratio of thermal to magnetic pressure, 
\begin{equation} \label{eq:beta}
    \beta \equiv \frac{8\pi p_g} {B^2} .
\end{equation}
The relevant frequencies are the gas sound frequency (with $c_s$ being the isothermal gas sound speed),
\begin{equation} \label{eq:ws}
    \omega_s \equiv k c_s ;
\end{equation}
the Alfv\'en \textit{and} CR-streaming frequency,
\begin{equation}\label{eq:wa}
\omega_{\rm A} \equiv  \bm{k \cdot \vrm_{\rm A}};
\end{equation}
the CR diffusion frequency,
\begin{equation}\label{eq:wd}
\omega_d \equiv \kappa \  (\bm{\hat{b} \cdot k})^2 ;
\end{equation}
the free-fall frequency,
\begin{equation}\label{eq:w_ff}
    \omega_{\rm ff} \equiv \frac{g}{c_s} = \frac{c_s}{ H},
\end{equation}
where $H$ is the gas scale height; the cooling frequency,
\begin{equation} \label{eq:w_c}
    \omega_c \equiv \frac{\rho^2 \Lambda}{p_g};
\end{equation}
the ion-ion collision frequency $\nu_{\rm ii}$ and the associated Braginskii viscous frequency,
\begin{equation} \label{eq:wb_def}
    \omega_{\rm B} \equiv \nub (\bm{\hat{b}\cdot k})^2 \approx \frac{p_g}{ \rho \nu_{\rm ii}} (\bm{\hat{b}\cdot k})^2;
\end{equation}
and the conductive frequency,
\begin{equation}
    \omega_{\rm cond} \equiv \chi_{\rm B} (\bm{\hat{b}\cdot k})^2,
\end{equation}
where $\chi_{\rm B} = \kappa_{\rm B} / n k_{\rm B}$ is the thermal diffusion coefficient. We can relate the diffusive timescales by defining the thermal Prandtl number,
\begin{equation} \label{eq:Pr}
    {\rm Pr} \equiv \frac{\nub}{\chi_{\rm B}}.
\end{equation}
We use ${\rm Pr} = 0.02$ as the default thermal Prandtl number in this work ($\approx$ square root of the electron-to-ion mass ratio), i.e. heat conduction due to electrons operates on a much shorter timescale than viscous forces due to the ions. We also define the ratio of the CR diffusion coefficient to the Braginskii viscosity,
\begin{equation} \label{eq:phi}
    \Phi \equiv \frac{\kappa}{\nub},
\end{equation}
which turns out to be an important parameter quantifying the suppression of the CRBI and CRAB instability by CR diffusion (see also \citealt{kqs2020}). 
\subsection{Validity of the model} 
\subsubsection{Thermal gas}\label{sec:validity_thermal_gas}
The CR entropy mode describes the response of the two-fluid CR--thermal gas system to a CR pressure perturbation. Since CRs propagate along field lines at the Alfv\'en speed, this mode is characterised by a frequency of order $\omega_{\rm A}$ (absent CR diffusion). The collisional (Braginskii MHD) regime then requires that $\omega_{\rm A} \ll \nu_{\rm ii}$, or equivalently $k l_{\rm mfp} \ll \beta^{1/2}$. In addition, the assumption that the heat fluxes are due to electrons (${\rm Pr} \approx 0.02$) is valid if the electron-ion thermal equilibration rate is $\gg \omega_{\rm A}$. The electron-ion thermal equilibration rate is slower than the ion-ion collision rate by the square root of the electron-to-ion mass ratio,  $\tau_{\rm eq}^{-1} \sim (m_e / m_i)^{1/2} \nu_{\rm ii}\sim \nu_{\rm ii}/40$, which means that ions and electrons are thermally coupled for $\omega \sim \omega_{\rm A}$ modes if $k l_{\rm mfp} \lesssim \beta^{1/2} / 40$. We will typically assume $\beta \sim 100$ and shall therefore consider $k l_{\rm mfp} \lesssim 1$. 

We further note that the expression for the pressure anisotropy in \eqref{eq:Deltap} ignores the effect of heat fluxes on $\Delta p$ (\citealt{mikhailovskii_tsypin_1971}). In particular, in the collisional limit the pressure anisotropy in the presence of heat fluxes is (e.g., \citealt{schekochihin_2010}),
\begin{equation} \label{eq:dp_q}
    \Delta p = 3 \rho \nub \Big( \bm{\hat{b} \hat{b} : \nabla \vrm} - \frac{1}{3} \bm{\nabla \cdot \vrm} - \frac{\bm{\nabla \cdot}[ (q_\perp - q_\parallel)\bm{\hat{b}}] + 3 q_\perp \bm{\nabla \cdot \hat{b}} }{3 p_\parallel} \Big),
\end{equation}
where $q_\perp$ and $q_\parallel$ are the parallel fluxes of perpendicular and parallel heat, respectively. In the high-collisionality limit with $\Delta p \ll p_g$, the heat fluxes are $q_\perp \approx q_\parallel / 3 \approx - \kappa_{\rm B} \mathbf{\hat{b} \cdot \nabla} T$ where $\kappa_{\rm B}$ is the thermal conductivity, $T = p_g / n k_{\rm B}$ and $p_g$ is given by \eqref{eq:ptot}. In this limit, we find that the results of this paper are not significantly affected by the heat-flux term in \eqref{eq:dp_q}. We thus use eq. \eqref{eq:Deltap} for simplicity throughout this work. However, we note that the collisional expressions for $\Delta p$, $q_\perp$ and $q_\parallel$ are not valid on scales approaching the ion mean free path. In this limit, separate evolution equations for the ion and electron perpendicular and parallel pressures should be considered. We briefly discuss this regime in Section \ref{sec:collisionless} and Appendix \ref{app:lf}, focusing on the ions. Our results suggest that the CRBI is qualitatively similar in the low-collisionality limit (Figure \ref{fig:lf}).

In the dilute and hot ICM, the ion-ion collision rate is small
\begin{equation} \label{eq:nuii_icm}
\nu_{\rm ii} \sim \frac{  n_{\rm i} e^4 \pi \ln \Lambda }{m_{\rm i}^{1/2} (k_{\rm B} T)^{3/2}} \sim 8 \times 10^{-14} \ {\rm s} ^{-1} \ \Big(\frac{T}{5 \times 10^7 \ \rm{K}} \Big)^{-3/2} \frac{n_{\rm i}}{0.01 \ {\rm cm}^{-3}},
\end{equation}
for a Coulomb logarithm $\ln \Lambda \approx 38$. The collision rate in \eqref{eq:nuii_icm} corresponds to a mean free path of order $0.1$ kpc.

\subsubsection{Cosmic rays}
The CR pressure equation (eq. \ref{eq:pc}) is a good model for the cosmic rays if the collision frequency of the energetically important GeV CRs is much larger than any other timescale of interest. As pointed out in \cite{kqs2020}, the GeV CR collision frequency (the rate at which the pitch angle changes by order unity, due to scattering by magnetic fluctuations) is likely much higher than the thermal ion-ion collision frequency in the ICM:
\begin{equation}
    \nu_{\rm CR} \sim \Omega_0 \Big(\frac{\delta B_\perp}{B} \Big)^2 \sim 10^{-8} \ {\rm s} ^{-1} \ \Big(\frac{\delta B_\perp / B}{10^{-3}} \Big)^2,
\end{equation}
where $\Omega_0$ is the non-relativistic gyro-frequency and $\delta B_\perp / B$ is evaluated for fluctuations whose wavelength parallel to the mean B-field is of order the Larmor radius of the GeV particles. The above collision frequency corresponds to a CR mean free path of order 1 pc, which approximately corresponds to the empirically derived CR mean free path in the Milky Way (e.g., \citealt{amato_blasi_18}). This suggests that treating cosmic rays as collisional on thermal-ion-mean-free-path scales is a reasonable model for the ICM.  

In the limit of good coupling between the GeV cosmic rays and the self-excited Alfv\'en waves (large pitch-angle scattering rate), CR transport is to leading order described by Alfv\'enic streaming. Damping of the self-excited Alfv\'en waves introduces corrections to Alfv\'enic streaming. We model this by also including CR diffusion in our linear analysis. As we will show, significant CR diffusion suppresses the CRBI on small scales. 

\subsection{Background equilibrium} \label{sec:equil}
We consider static background equilibria with constant $\bm{B} = (B_x, 0, B_z)$, in which the CR and gas pressure gradients balance gravity, $\bm{g}=-g\bm{\hat{z}}$,
 \begin{equation} \label{eq:back_general_HE}
     \frac{d}{dz} (p_g + p_c) = - \rho g.
 \end{equation}
If CR diffusion does not affect the background state (either because $\kappa = 0$ or $dp_c/d z={\rm const}$), CRs are in equilibrium according to \eqref{eq:pc} if $p_c \propto {\rm v_A}^{-4/3}$, or equivalently,
\begin{equation} \label{eq:back_general_pc}
    p_c \propto \rho^{2/3}.
\end{equation}
In this work we perturb equilibria that satisfy \eqref{eq:back_general_HE} and \eqref{eq:back_general_pc}, although these equations do not yet constrain the background temperature $T(z)$. We will consider different temperature profiles in Sections \ref{sec:icm} and \ref{sec:discussion}.

\subsection{Linearised equations}\label{sec:lin_eq}
We carry out a linear stability calculation of the CR--thermal gas equations (Section \ref{sec:eqns}) for backgrounds that satisfy \eqref{eq:back_general_HE} and \eqref{eq:back_general_pc}. We focus on short-wavelength modes satisfying $kH \gg 1$ and perform a local WKB calculation in which all perturbed quantities are assumed to vary as $\delta X(\bm{r}, t) \propto \exp \Big(i \bm{k \cdot r} - i \omega t \Big)$. We consider two coordinate systems in this work. We will mostly use the usual cartesian coordinate system with the $z$-axis anti-parallel to $\bm{g}$ and $x$ and $y$ defining the plane perpendicular to gravity. We define associated polar angles such that $B_x = B \sin \theta_{\rm B}$,  $B_z = B \cos \theta_{\rm B}$, $k_x = k \sin \theta_k \cos \phi_k$, $k_y = k \sin \theta_k \sin \phi_k$ and $k_z = k \cos \theta_k$. We use $\phi_k = 0$ in all the figures except Figure \ref{fig:angle}, because $\phi_k=0$ captures the fastest-growing mode. For analytic purposes, it is convenient to also use a coordinate system aligned with the direction of the magnetic field. We define $\perp$ and $\parallel$ to denote the directions perpendicular and parallel to the magnetic field. We will use this coordinate system in Section \ref{sec:grav} when we derive an approximate dispersion relation, as it makes the analytics more tractable. However, when we plot numerically calculated growth rates of the instability in Figures \ref{fig:simple_picture}--\ref{fig:lf}, we adopt the more standard notation with directions/angles defined relative to the positive z-direction, i.e. $\cos\theta_{B} \equiv - \bm{\hat{b}\cdot \hat{g}}$ and $\cos\theta_{k} \equiv  -\bm{\hat{k}\cdot \hat{g}}$. 

The linearised equations  are
\begin{equation} \label{eq:drho}
    \omega \frac{\delta \rho}{\rho} = \bm{k \cdot {\rm v}} - i \vrm_z \frac{d \ln \rho}{d z} \times {\rm BG}, 
\end{equation}

\begin{gather}
\begin{aligned}\label{eq:dv}
    \omega \bm{{\rm v}} = & \bm{k} c_s^2 \Big(  \frac{\delta p_\perp}{p_g} + \eta \frac{\delta p_c}{p_c} + \frac{{\rm v_A}^2}{c_s^2} \frac{\delta B_\parallel}{B} \Big) - \omega_{\rm A} {\rm v_A} \frac{\bm{\delta B}}{B} \\ & - \bm{\hat{b}}(\bm{\hat{b}\cdot k})  c_s^2 \frac{\delta \Delta p}{p_g} + i \bm{g} \frac{\delta \rho}{\rho},
\end{aligned}
\end{gather}

\begin{equation} \label{eq:dB_x}
    \omega \frac{\bm{\delta B_\perp}}{B} = - k_\parallel \bm{{\rm v}_\perp},
\end{equation}
\begin{equation} \label{eq:dB_z}
    \omega \frac{\delta B_\parallel}{B} =  \bm{k_\perp \cdot  {\rm v}_\perp },
\end{equation}
\begin{gather}
\begin{aligned} \label{eq:dpg}
    \omega \frac{\delta p_g}{p_g} = &  \gamma \bm{k \cdot {\rm v}} - i(\gamma-1) \omega_{\rm cond} \Big( \frac{\delta p_g}{p_g} - \frac{\delta \rho}{\rho} \Big)  + \chi \eta (\gamma -1) \omega_{\rm A} \frac{\delta p_c}{p_c} \\ & 
    - i (\gamma - 1) \omega_c \frac{\partial \ln \Lambda}{\partial \ln T} \frac{\delta p_g}{p_g} - i (\gamma - 1) \omega_c \Big(2-\frac{\partial \ln \Lambda}{\partial \ln T} \Big) \frac{\delta \rho}{\rho} \\ & + {\rm BG} \times \Big[ -i \vrm_z \frac{d \ln p_g}{dz}  - i \chi \eta (\gamma -1) \delta {\rm v_{A,}}_z \frac{d \ln p_c}{dz} \\ & +  \frac{i (\gamma -1)}{T} \bm{\nabla \cdot} (\chi_{\rm B} \mathbf{ \boldsymbol\delta \hat{b} \hat{b} \cdot \boldsymbol\nabla}T + \chi_{\rm B} \mathbf{ \hat{b} \boldsymbol\delta \hat{b} \cdot \boldsymbol\nabla}T) \Big], 
\end{aligned}
\end{gather}
\begin{equation} \label{eq:dDp_Brag}
    \frac{\delta \Delta p}{p_g} = i \frac{\rho \nub}{p_g} (2 k_{\parallel} {\rm v}_\parallel -\bm{ k_\perp \cdot {\rm v}_\perp}) .
\end{equation}
\begin{gather}
\begin{aligned} \label{eq:dpc}
    \omega \frac{\delta p_c}{p_c} = & \frac{4}{3} \bm{k \cdot {\rm v}} - \frac{2}{3} \chi \omega_{\rm A} \frac{\delta \rho}{\rho}  + \frac{1}{3}\beta  \chi \omega_{\rm A} \frac{\delta \Delta p}{p_g}  + (\chi \omega_{\rm A} - i \omega_d) \frac{\delta p_c}{p_c} \\ & + {\rm BG} \times \Big[ - i {\rm v}_z \frac{d \ln p_c}{dz} + \frac{2}{3}i  \chi{\rm v_{A,z}} \frac{d \ln \rho}{dz} \Big(\frac{\delta p_c}{p_c} - \frac{\delta \rho}{\rho} \Big)  \\ &  +  \frac{i}{p_c} \mathbf{\nabla \cdot} \Big( \kappa \mathbf{\boldsymbol\delta \hat{b} \hat{b} \cdot \boldsymbol\nabla} p_c + \kappa  \mathbf{ \hat{b} \boldsymbol\delta \hat{b} \cdot \boldsymbol\nabla}p_c  \Big) \Big] , 
\end{aligned}
\end{gather}
where ${\rm BG}$ in equations \eqref{eq:drho}--\eqref{eq:dpc} is a flag equal to 1 or 0. It specifies whether gradients from the background equilibrium in \eqref{eq:back_general_HE} and \eqref{eq:back_general_pc}, i.e. terms of the form ${\rm v}_z d\rho / dz$,  are included in the calculation. Without the background gradients equilibrium is not satisfied and so the BG terms should formally be kept in the linear perturbation analysis. However, for analytic simplicity, we set ${\rm BG} = 0$ when we discuss the physics of the instability and derive approximate dispersion relations in Sections \ref{sec:pres_bal}, \ref{sec:gravity_simple} and \ref{sec:grav}. This turns out to be a reasonable approximation, as the CRBI is not directly driven by the background gradients. We do, however, include background gradients, i.e. we set ${\rm BG} = 1$, when we solve for the exact eigenmodes numerically (Section \ref{sec:icm} and all the figures presented in this paper).

\section{Physical overview of the instability} \label{sec:prelims}
In this section, we provide an overview of the key physics describing the CRBI. To start, we give a brief summary of entropy and gravity waves in stratified CR MHD (\ref{sec:modes}). This will set the stage for our analysis and elucidate how the CRBI is related to other instabilities that may operate in the ICM (such as thermal instability or other buoyancy instabilities).

\subsection{Gravity and entropy modes in stratified media} \label{sec:modes}

 \begin{figure}
  \centering
    \includegraphics[width=0.45\textwidth]{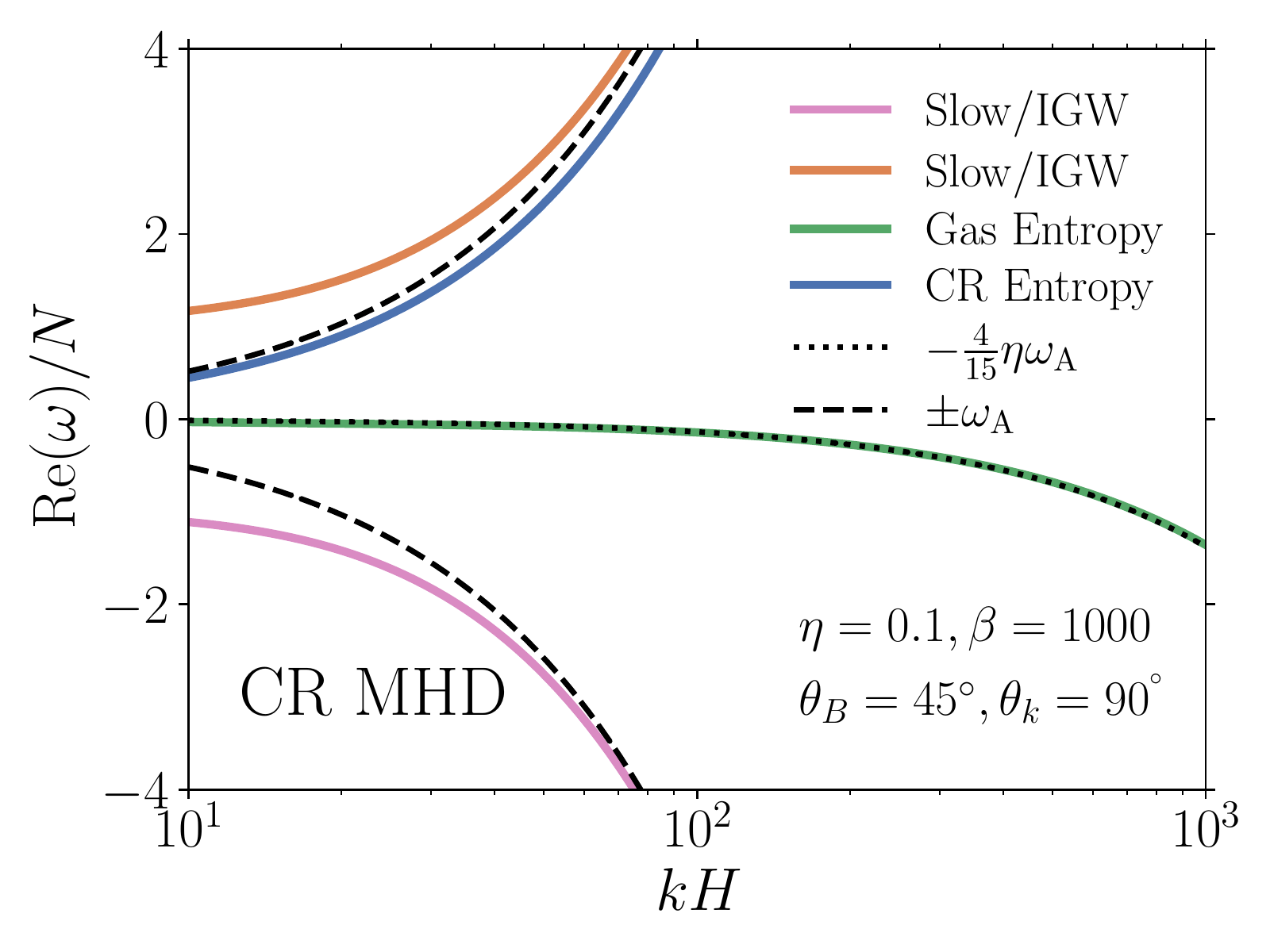}

  \caption[Oscillation frequencies of gravity/slow and entropy modes in stratified, collisional CR MHD.]{Oscillation frequencies of gravity/slow and entropy modes in stratified, collisional CR MHD. The frequencies are normalised by the  Brunt–V\"{a}is\"{a}l\"{a} frequency. Here and in other figures (except Figure \ref{fig:angle}), we use $\phi_k = 0$. The two modes that are characterised by the buoyancy frequency at long wavelengths (orange and pink lines) become the MHD slow magnetosonic modes at short wavelengths. The oscillation frequency of the gas-entropy mode (green line) is due to CR heating. The blue line shows the CR-entropy  mode with frequency $\approx \omega_{\rm A}$, with the small deviation arising from finite $\eta$. We will show that gravity destabilises the CR-entropy mode in low-collisionality MHD. We term this instability the CR buoyancy instability (CRBI).  \label{fig:all_modes_real_crmhd}}
\end{figure}

 \begin{figure}
  \centering
    \includegraphics[width=0.45\textwidth]{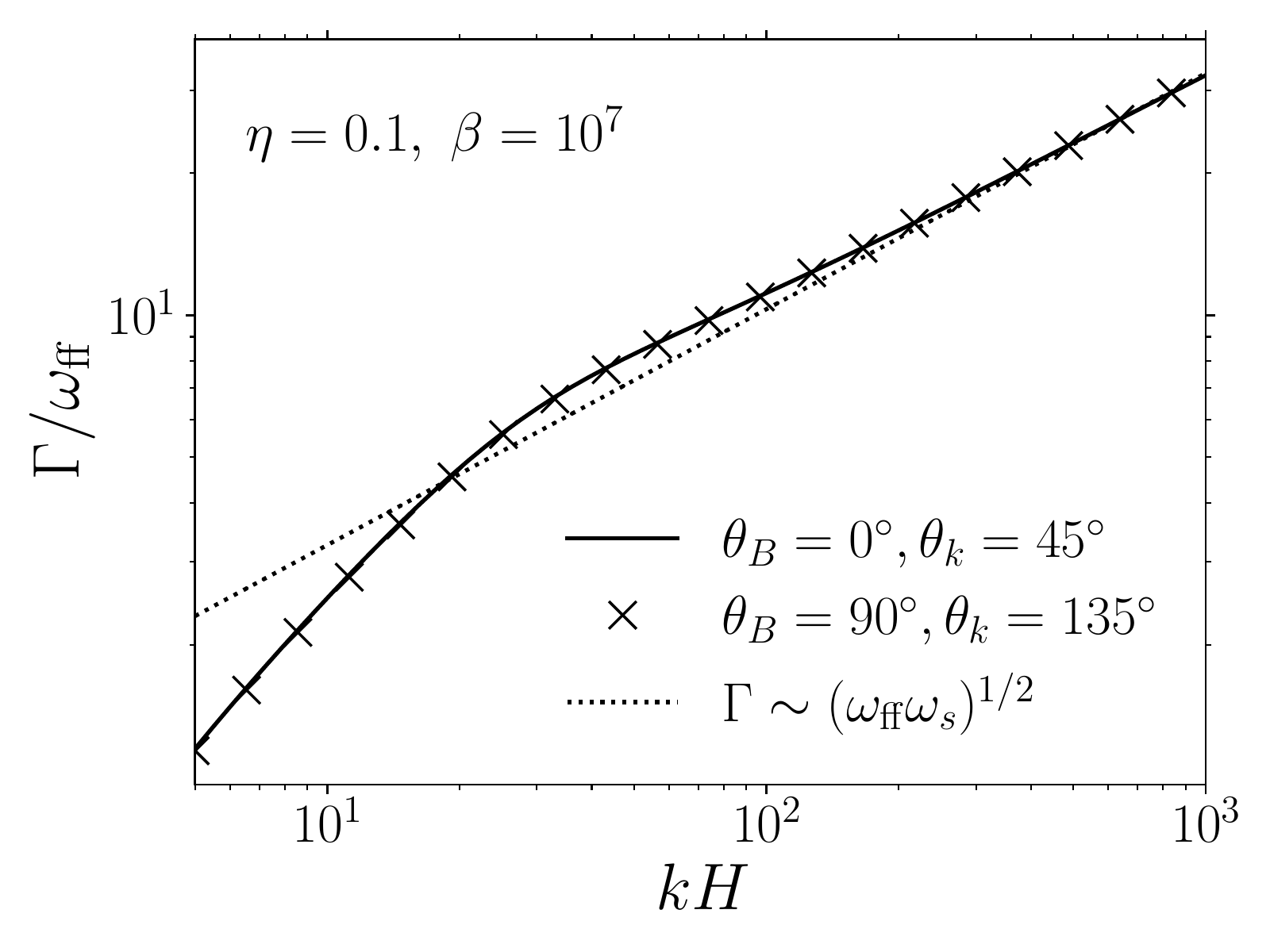}
  \caption[Growth rates of the CRBI in the high-$\beta$ regime.]{Growth rates of the CRBI for $\beta=10^7$, $\eta=0.1$, and $H=1000 l_{\rm mfp}$. The angles $\theta_B$ and $\theta_k$ are the directions of $\bm{B}$ and $\bm{k}$ with respect to the positive $z$-direction ($\bm{g}=-g\bm{\hat{z}}$). At high $\beta$, damping by anisotropic pressure is negligible over a range of $k$ and the instability is well described by the simple model in Section \ref{sec:gravity_simple}. The dotted line labelled $\sim \sqrt{\omega_s \omega_{\rm ff}}$ shows the predicted growth rate from  eq. \eqref{eq:simple_LF}. This elucidates the physics of the CRBI in its simplest form. Figures \ref{fig:mode} -- \ref{fig:lf} show results for more realistic ICM conditions. \label{fig:simple_picture}}
\end{figure}

\subsubsection{MHD modes} 
In a gravitationally stratified medium without magnetic fields, the hydrodynamic entropy mode becomes an internal gravity wave characterised by the Brunt–V\"{a}is\"{a}l\"{a} frequency, 
\begin{equation}\label{eq:N2}
 N^2 = \frac{g}{\gamma} \Big( \frac{d \ln p_g}{dz} - \gamma \frac{d \ln \rho}{dz} \Big).
\end{equation} 
In MHD, gravitational stratification affects the MHD slow magnetosonic modes, as both buoyancy and magnetic tension act as the modes' restoring forces. The resulting mode resembles the hydro IGW at long wavelengths, and the standard MHD slow mode (absent gravity) at short wavelengths. In contrast to hydrodynamics, in MHD there is also a mode that is unaffected by buoyancy despite a non-zero density perturbation. We refer to this mode as the MHD entropy mode. 

We now compute these modes more explicitly in high-$\beta$ collisional MHD, first assuming that no cosmic rays are present. For simplicity, we shall consider the $2$D case with $\bm{B} = B \bm{\hat{z}}$, $\bm{g}=-g\bm{\hat{z}}$ and $\bm{k}$ in the $x-z$ plane. We start with the momentum equation, 
\begin{equation} \label{eq:strat_momentum}
    \omega \bm{{\rm v}} = \bm{k} c_s^2 \Big(\frac{\delta p_g}{p_g} +  \frac{{\rm v_A}^2}{c_s^2} \frac{\delta B}{B} \Big) - \omega_{\rm A} {\rm v_A} \frac{\bm{\delta B}}{B} - i \frac{\delta \rho}{\rho} g \bm{\hat{z}}.
\end{equation}
Crossing the momentum equation twice with $\bm{k}$ and taking the z-component we find,
\begin{equation}
    \omega^2 (k_z k_x {\rm v}_x - k_x^2 {\rm v}_z ) = \omega_{\rm A} {\rm v_A} k^2 k_x {\rm v}_x + i \omega g k_x^2 \frac{\delta \rho}{\rho}.
\end{equation}
In the Boussinesq limit ($\delta p_g / p_g \ll \delta \rho / \rho$), the adiabatic gas entropy equation $ds / dt = 0$ (eq. \ref{eq:dpg} without CRs, cooling and conduction) implies that 

\begin{equation}
    \omega \frac{\delta \rho}{\rho} = \frac{i}{\gamma} \vrm_z \Big(\frac{d \ln p_g}{dz} - \gamma \frac{d \ln \rho}{dz}    \Big) = i \frac{\rm v_z}{g} N^2.
\end{equation}
Using incompressibility $k_x {\rm v}_x \approx - k_z {\rm v}_z$ we find the dispersion relation for MHD slow modes modified by gravity, 
\begin{equation}
    \omega^2 k^2   = \omega_{\rm A} {\rm v_A} k^2 k_z + k_x^2\frac{g}{\gamma} \Big( \frac{d \ln p_g}{dz} - \gamma \frac{d \ln \rho}{dz} \Big), 
\end{equation}
or,
\begin{equation}
    \omega^2   = \omega_{\rm A}^2 +  \frac{k_x^2}{k^2} N^2.
\end{equation}
For subdominant magnetic tension (first term on RHS) we get the usual dispersion relation for the two hydrodynamic IGWs, $\omega = \pm N k_x / k$. At short wavelengths where magnetic tension dominates, the two gravity waves satisfy the standard dispersion relation for the MHD slow modes at high $\beta$.

Unlike the hydrodynamic entropy mode, the MHD entropy mode does not pick up a buoyancy response. To see why this is the case, we first note that a mode with $\omega=0$ and $\mathbf{\vrm}=0$, but finite $\delta \rho/\rho$, satisfies the continuity (eq. \ref{eq:drho}), induction (eq. \ref{eq:dB_x}--\ref{eq:dB_z}), as well as the gas-entropy (eq. \ref{eq:dpg} absent heating, cooling and conduction) equations. In hydrodynamics, however, this mode does not satisfy the momentum equation (see eq.  \ref{eq:strat_momentum} without the perturbed magnetic field), as the direction of gravity and $\bm{k}$ are generally not co-linear. By contrast, in MHD the perturbed pressure and magnetic-tension terms (first and second terms on the RHS of eq. \ref{eq:strat_momentum}) are mutually orthogonal and can exactly cancel the perturbed gravitational force. As a result, while a mode with $\omega=0$ and $\mathbf{\vrm}=0$, but finite $\delta \rho/\rho$, is not an eigenmode in stratified hydrodynamics, it is an eigenmode in stratified MHD and involves a finite $\bm{\delta B}$ perturbation. When magnetic tension is negligible, i.e. in the hydrodynamic limit with $\omega_{\rm A} \ll N$ (long wavelengths or $\beta \rightarrow \infty$), the $\omega=0$ mode can only be satisfied if $\delta B / B \gg \delta \rho / \rho$. The hydrodynamic variables are therefore essentially unperturbed in this limit, which is consistent with the result in stratified hydrodynamics.   

\subsubsection{CR MHD modes}
When Alfv\'enically streaming CRs are present, the CR pressure equation (eq. \ref{eq:pc}) introduces a new mode, which we refer to as the CR entropy mode. Because CRs are assumed to stream at the Alfv\'en speed along field lines, the CR entropy mode is characterised by the Alfv\'en frequency, $\omega = \chi \bm{k \cdot {\rm v_A}}=\chi \omega_{\rm A}$ (although the eigenfrequency can deviate appreciably from the Alfv\'en frequency if CR diffusion is important or if $\eta \gtrsim 1$ due to the fact that the CR entropy mode is then associated with significant density fluctuations). The $\chi$ factor in front of $\omega_{\rm A}$ reflects the fact that the CR entropy mode propagates down the CR pressure gradient. For $\eta \rightarrow 0$ the impact of CRs on the thermal gas is small, and so the CR entropy mode does not perturb the thermal gas. Using that $2 d \ln \rho / dz = 3 d\ln p_c/dz$ in equilibrium, \eqref{eq:dpc} becomes,
\begin{equation}\label{eq:pc_en}
    \Big( \omega - \chi \omega_{\rm A} + i \omega_d -i \chi {\rm v}_{{\rm A},z} \frac{d \ln p_c}{dz} \Big) \frac{\delta p_c}{p_c} \approx 0,
\end{equation}
with solution, 
\begin{equation}\label{eq:pc_en2}
    \omega = \chi \omega_{\rm A} - i \omega_d +i \chi {\rm v}_{{\rm A},z} \frac{ d \ln p_c}{dz}.
\end{equation}
Diffusive corrections to CR streaming act to damp the mode. The CR background gradient term also introduces an imaginary part that looks like damping. However, the more accurate interpretation of this term is that as the mode propagates down the CR pressure gradient, the perturbation amplitude normalized by the local CR pressure, $\delta p_c / p_c(z)$, remains constant. 

Because streaming CRs heat the gas at a rate $- \chi \bm{ {\rm v_A} \cdot \nabla} p_c$, they also modify the gas-entropy mode by giving it a real (oscillatory) frequency, which at small $\eta$ is $\omega \approx - 4\eta \omega_{\rm A}/15 $ (\citealt{kq2020}). Importantly, to leading order CR heating does not significantly affect the growth/damping of the gas-entropy mode (e.g., due to thermal instability), just its real frequency.

We summarise the discussion above by showing the oscillation frequencies of gravity and entropy modes in stratified, collisional CR MHD in Figure \ref{fig:all_modes_real_crmhd}. The frequencies are normalised by the  Brunt–V\"{a}is\"{a}l\"{a} frequency. The two modes that are characterised by the buoyancy frequency at long wavelengths (orange and pink lines) become the MHD slow magnetosonic modes at short wavelengths. The gas-entropy mode (green line) is unaffected by buoyancy, and its oscillation frequency is due to CR heating. The blue line shows the CR-entropy  mode with frequency $\approx \omega_{\rm A}$. Figure \ref{fig:all_modes_real_crmhd} suggests that the CR-entropy mode is not significantly affected by buoyancy in collisional MHD. However, we will show that gravity destabilises the mode in low-collisionality MHD (i.e., on small scales), which we term the CR buoyancy instability (CRBI). 

\subsection{Compressible CR entropy mode due to streaming} \label{sec:pres_bal}
Surprisingly, the CR entropy mode becomes highly compressible at short wavelengths due to the influence of a finite mean free path in the background plasma, i.e. due to $\Delta p$. At high $\beta$, pressure balance implies $\delta p_c + \delta p_\perp \approx 0$. In the isothermal limit due to rapid conduction ($\rm{Pr} \ll 1$), and neglecting CR diffusion, this can be rewritten as (see eq. \ref{eq:dpc}):
\begin{equation} \label{eq:pbal}
    \frac{\delta \rho}{\rho} + \eta \frac{\frac{4}{3} \bm{k \cdot {\rm v}} - \frac{2}{3} \chi \omega_{\rm A} \frac{\delta \rho}{\rho} + \frac{1}{3} \beta \omega_{\rm A} \chi \frac{\delta \Delta p}{p_g}}{\omega-\chi \omega_{\rm A} } \approx 0.
\end{equation}
In the limit of long wavelengths/high collisionality such that $\delta \Delta p$ is negligible, pressure balance is achieved if $\delta \rho / \rho \rightarrow 0$, i.e. $k_x {\rm v}_x + k_y {\rm v}_y  + k_z {\rm v}_z = 0$ or $\bm{k_\perp \cdot {\rm v}_\perp} + k_\parallel {\rm v}_\parallel = 0$. This is the standard result that pressure-balanced modes are nearly incompressible at high $\beta$. 

At shorter wavelengths and high $\beta$, the $\delta \Delta p / p_g$ term is dominant for the CR entropy mode with $\omega \approx \omega_{\rm A}$. Pressure balance then requires that the pressure anisotropy is minimised, $\delta \Delta p \approx 0$. This condition leads to the unusual requirement that the pressure-balanced mode is highly compressible. In Braginskii MHD $\Delta p \propto (2 k_\parallel {\rm v}_\parallel - \bm{k_\perp \cdot  {\rm v}_\perp)}$ so that \eqref{eq:pbal} is satisfied when,
\begin{equation} \label{eq:compressible_brag}
    \bm{k_\perp \cdot {\rm v}_\perp} \approx 2 k_\parallel {\rm v}_\parallel.
\end{equation}
The exact relation satisfied by $\vrm_\perp$ and $\vrm_\parallel$ can be different in the collisionless regime, i.e. below the ion mean free path, where Braginskii MHD is no longer valid. However, the qualitative conclusion remains the same and the mode is very compressible. We next show how the pressure-balanced compressible mode is destabilised by gravity. 

\subsection{How gravity destabilises the compressible  CR entropy mode} \label{sec:gravity_simple}

 To show how gravity destabilises the compressible CR entropy mode discussed above, we consider a simple model in which we ignore magnetic tension and damping by anisotropic pressure (viscosity). As in Section \ref{sec:modes}, we consider the case $\bm{B} = B\bm{\hat{z}}$,  $\bm{g} = - g\bm{\hat{z}}$, and consider a mode with $k_x {\rm v}_x = \alpha k_z {\rm v}_z$ imposed by pressure balance. As described in Section \ref{sec:pres_bal}, $\alpha=-1$ in standard MHD, while $\alpha=2$ in Braginskii MHD with CRs. The momentum equations are,
\begin{equation}\label{eq:grav_simple_vx}
    \omega {\rm v}_x = \frac{k_x}{\rho}\delta P_{\rm tot},
\end{equation}
\begin{equation}\label{eq:grav_simple_vz}
    \omega {\rm v}_z = \frac{k_z}{\rho} \delta P_{\rm tot} - i \omega_{\rm ff} c_s \frac{(1+\alpha) k_z {\rm v}_z}{\omega},
\end{equation}
where in the ${\rm v}_z$ equation we used that $\omega \delta \rho / \rho \approx (1+\alpha) k_z {\rm v}_z$. Multiplying \eqref{eq:grav_simple_vx} by $k_x$ and subtracting from  \eqref{eq:grav_simple_vz} times $\alpha k_z$, we find that
\begin{equation}
    \frac{\delta P_{\rm tot}}{\rho} (\alpha k_z^2 - k_x^2) = i \alpha (1+\alpha) k_z c_s \omega_{\rm ff} \frac{k_z {\rm v}_z}{\omega} .
\end{equation}
Using this expression for $\delta P_{\rm tot}$ back in equations \eqref{eq:grav_simple_vx} and \eqref{eq:grav_simple_vz} gives a simple dispersion relation:
\begin{equation}\label{eq:simple_LF}
       \omega  \approx  \frac{1}{\sqrt{2}}(1+i) \Big[ (1+\alpha) k_z c_s \omega_{\rm ff} \frac{kx^2}{ \alpha k_z^2 - k_x^2}  \Big]^{1/2} \sim \sqrt{\omega_s \omega_{\rm ff}}.
\end{equation}
In this simplified picture gravity leads to growth rates that are of order $\sim \sqrt{\omega_{\rm ff}\omega_s} \gg \omega_{\rm ff}$. The above analysis can be easily repeated for the case $\bm{g} \perp \bm{B}$ (i.e. horizontal magnetic field in a vertical gravitational field), with very similar results. We stress that the scaling $\sim \sqrt{\omega_{\rm ff}\omega_s}$ in \eqref{eq:simple_LF} does not describe compressible sound waves, which are approximately longitudinal at high $\beta$: $\alpha \approx k_x^2 / k_z^2$ and the denominator in \eqref{eq:simple_LF} is approximately zero.   

Akin to standard buoyancy instabilities, the mode found here is destabilised by the unbalanced gravitational force acting on the mode's density fluctuations. However, in contrast to standard buoyancy instabilities, such as thermal convection in stars or the magneto-thermal instability (MTI; \citealt{mti_balbus}) and the heat-flux-driven buoyancy instability (HBI; \citealt{hbi}) in clusters, the density fluctuations in the CR-driven instability are not due to the background stratification of the plasma. The density fluctuations at short wavelengths are instead due to CR streaming,  independent of the background stratification.

In subsequent sections we show that while magnetic tension does not affect the growth rate significantly, the effect of damping by anisotropic pressure should generally be retained. However, at sufficiently high $\beta$ there is a range of scales for which the simple model considered in this section provides a good picture of the instability (see eq. \ref{eq:no_dp_condition} below), and the growth rates are indeed $\sim \sqrt{\omega_s \omega_{\rm ff}}$. This regime is shown in Figure \ref{fig:simple_picture}. The solid black line shows the instability growth rate, computed from numerical solutions of the full set of linear equations in \ref{sec:lin_eq}, as a function of wavenumber for $\beta=10^7$, $\eta=0.1$, $\bm{B}$ antiparallel to $\bm{g}$ and $k_x = k_z$. The 'x' markers show the growth rate for $\bm{B}$ perpendicular to $\bm{g}$ and $k_x =- k_z$. The dotted line labelled $\sim \sqrt{\omega_s \omega_{\rm ff}}$ shows the predicted growth rate from \eqref{eq:simple_LF}, which matches the exact solution very well over a wide range in $k$.

We now discuss the instability in the astrophysically more relevant regimes where \eqref{eq:simple_LF} is not as accurate.

 \begin{figure}
  \centering
    \includegraphics[width=0.42\textwidth]{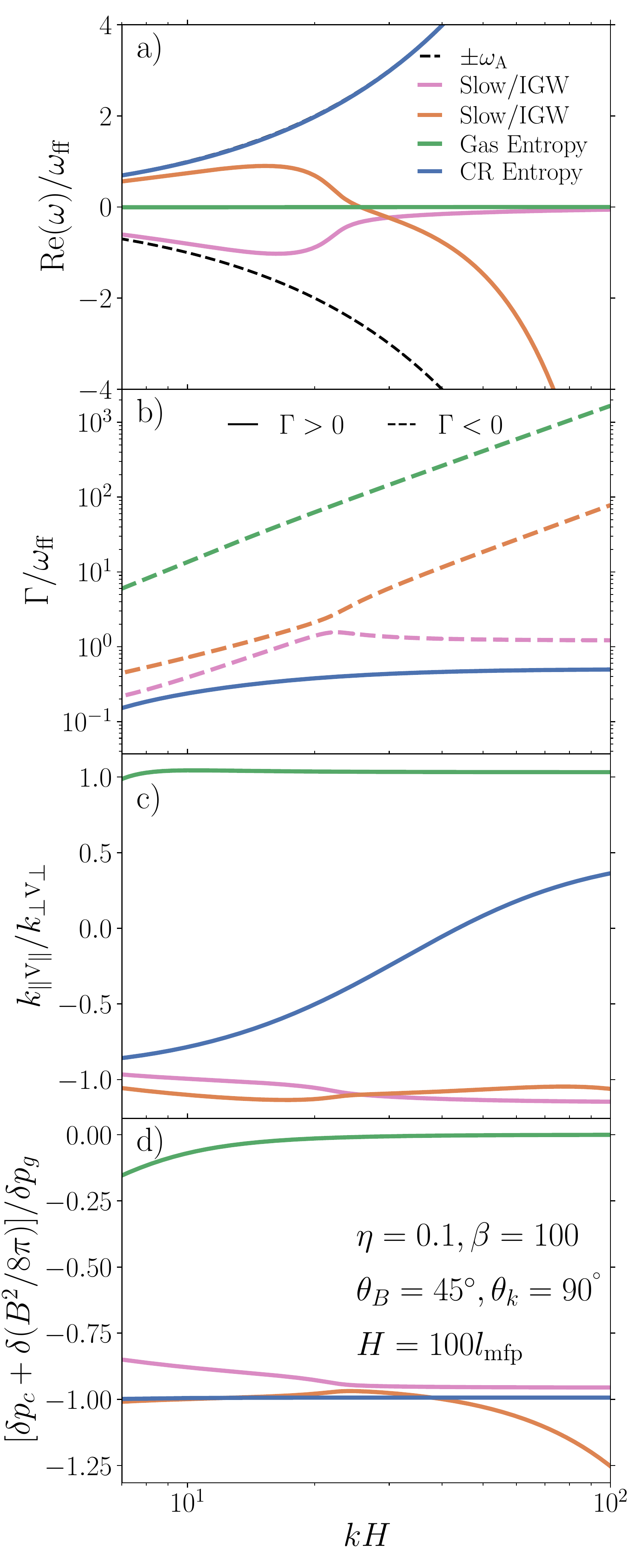}
  \caption[Properties of gravity and entropy modes in low-collisionality CR MHD.]{Mode properties for $\eta=0.1$, $\beta=100$ and $H=100l_{\rm mfp}$. $\theta_B=45^\circ$ and $\theta_k=90^\circ$ are the directions of $\bm{B}$ and $\bm{k}$ with respect to $\bm{\hat{z}}$ ($\bm{g}=-g\bm{\hat{z}}$), while $\perp$ and $\parallel$ are defined w.r.t. $\bm{B}$. We plot oscillation frequencies in panel a), growth rates in b) ($\Gamma >0$ corresponds to growth),  $k_\parallel \vrm_\parallel / k_\perp \vrm_\perp$, which quantifies the compressibility of the mode, in c), and $[\delta p_c + \delta (B^2/8\pi)] / \delta p_g$, which quantifies the degree of pressure balance, in d). The blue line shows the unstable CR entropy mode, the orange and pink lines are the MHD slow modes and the green line is the gas-entropy mode. All modes except the CR entropy mode are damped by low-collisionality effects. Long-wavelength CR entropy modes are approximately incompressible and the growth rates are small.  High-$k$ CR entropy modes become compressible due to CR streaming while maintaining pressure balance, and the growth rate reaches the plateau given by eq. \eqref{eq:gamma}. \label{fig:mode}}
\end{figure}

 \begin{figure}
  \centering
    \includegraphics[width=0.45\textwidth]{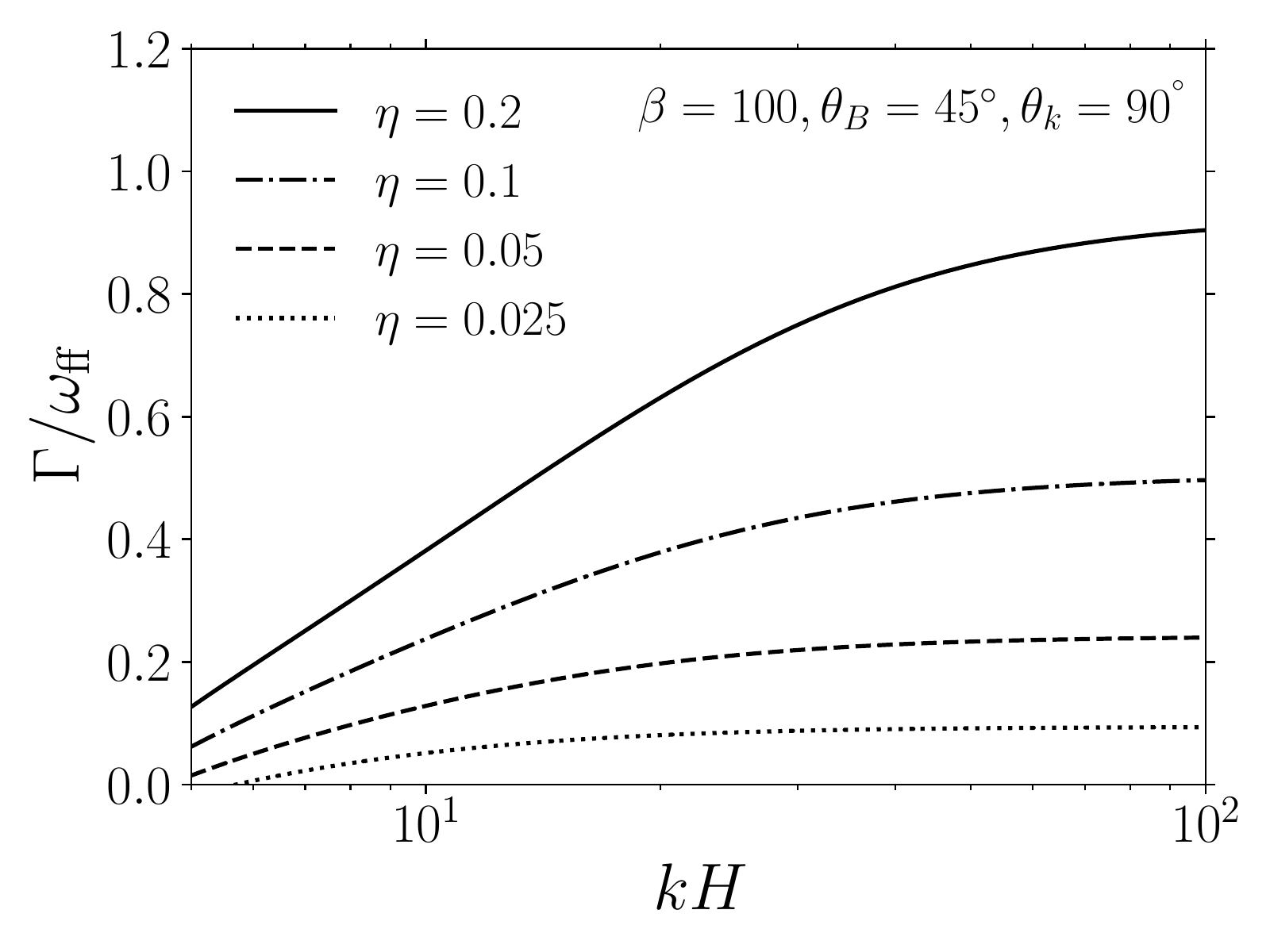}
  \caption[Growth rates of the CRBI for different CR pressure fractions.]{Growth rates of the CRBI for $\beta= 100$, $H = 100 l_{\rm mfp}$ and different ratios of CR pressure to gas pressure, $\eta$. The other parameters are set to the fiducial values (Section \ref{sec:icm}). The angles $\theta_B=45^\circ$ and $\theta_k=90^\circ$ are the directions of $\bm{B}$ and $\bm{k}$ with respect to the positive $z$-direction ($\bm{g}=-g\bm{\hat{z}}$). Even at very small CR pressures (small $\eta$), the instability still exists, but with reduced growth rates (eq. \ref{eq:gamma}). \label{fig:growth_vs_eta}}
\end{figure}

\section{Dispersion Relation and Growth Rates of Short-Wavelength Modes} \label{sec:grav}

In \ref{sec:gravity_simple} we neglected damping by anisotropic pressure (viscosity), which generally changes the growth rates relative to those predicted in \eqref{eq:simple_LF} (except for certain asymptotic limits, e.g. very large $\beta$ as in Figure \ref{fig:simple_picture}). In this section we present a more accurate calculation and derive an approximate dispersion relation. 

For simplicity, we here ignore background gradient terms (i.e. we set ${\rm BG=0}$  in equations \ref{eq:drho}--\ref{eq:dpc}) in our analytic derivation of the growth rates. Ignoring background gradients is a reasonable simplification because we find that short-wavelength modes (which are the fastest growing modes) are not significantly affected by explicitly including background gradients (even though they should formally be included). This is because the CRBI is due to streaming-induced compressibility and not background stratification.\footnote{However, we note that a nonzero CR pressure gradient along the magnetic field is generally necessary for equilibrium and to couple CRs to the thermal gas.} This is in contrast to thermal convection in stars or the MTI/HBI in clusters, which are driven by heat conduction and background temperature gradients. Our analytic results from this section will be supplemented by numerical solutions of the full system of linearised equations (equations \ref{eq:drho}-\ref{eq:dpc}) including background gradients in Section \ref{sec:icm}.

\subsection{Dispersion relation}\label{sec:disp}
In this section, it is convenient to work in a coordinate system aligned with the magnetic field: $\bm{B}=(0,0,B)$, $\bm{k} = (k_\perp, 0, k_\parallel)$ and $\bm{g} = (g_{\perp;1}, g_{\perp;2}, g_\parallel)$. In this coordinate system, and if we neglect background-gradient terms (which we do here), Alfv\'enic fluctuations of the form $\bm{\delta {B}}=(0,\delta B ,0)$ and $\bm{\delta {\rm v}}=(0,\delta {\rm v},0)$ decouple, which leaves velocity/magnetic field fluctuations in the $\bm{k}-\bm{B}$ plane for the remaining modes. If background gradients are kept, this is strictly true only if gravity is coplanar with $\bm{k}$ and $\bm{B}$. In the high-$\beta$ and isothermal limit ($\omega_{\rm cond} \gg \omega$), the third order dispersion relation for the slow-magnetosonic and CR-entropy modes can be found by crossing the momentum equation \eqref{eq:dv} twice with $\bm{k}$, taking the component parallel to $\bm{B}$ and using \eqref{eq:pbal}:
\begin{gather}
\begin{aligned} \label{eq:disp_slow}
   0 & =   \omega^2 \Big( \frac{2}{3} i \chi \eta \frac{\omega_B}{\omega_{\rm A}} \omega (2 k_\parallel^2 - k_\perp^2)+ k^2 \Delta \Big) + 3i k_\perp^2 \omega_B \omega \Delta  \\ & + 2 \chi \eta \frac{\omega_B}{\omega_{\rm A}} k_\parallel \omega_{\rm ff} c_s \omega \Big( -k^2  (\bm{\hat{b}\cdot \hat{g}}) + (\bm{k \cdot \hat{g}})k_\parallel \Big)   \\ &  -
    \omega_{\rm A} {\rm v_A} k^2 k_\parallel \Big( \frac{4}{3}i\eta \chi \frac{\omega_B}{\omega_{\rm A}} \omega + \Delta  \Big),
\end{aligned}
\end{gather}
where,
\begin{equation}
    \Delta = \omega - \chi \omega_{\rm A} + i \omega_d + \frac{4}{3}\eta \omega - \frac{2}{3}\eta \chi \omega_{\rm A}.
\end{equation}
The 3 solutions of the cubic dispersion relation in \eqref{eq:disp_slow} are the two slow magnetosonic waves and the CR entropy mode. The gas entropy mode is not present because our calculation assumed that modes are isothermal, $\delta p_g /p_g = \delta \rho / \rho$, which eliminates the gas entropy mode (i.e. the gas entropy equation acts as a constraint). The first term in \eqref{eq:disp_slow} comes from crossing the velocity perturbation twice with $\bm{k}$, the second term is the damping by Braginskii viscosity, the third term is the gravitational force and the last term comes from the perturbed magnetic tension. We note that the gravitational force term, which drives the instability, is zero if $\bm{g}$ is perpendicular to the $\bm{k}-\bm{B}$ plane. 

As explained in Section \ref{sec:gravity_simple}, the instability is driven by gravity mediated by compressibility induced by CR streaming. The compressibility is a consequence of the perturbed pressure anisotropy $\delta \Delta p$, characterised by the Braginskii viscous frequency $\omega_{\rm B}$, in the CR entropy equation. This suggests that compressibility effects are most important in the short-wavelength limit $\omega_B / \omega_{\rm A} \gg 1$. In this limit, the dispersion relation can be simplified to
\begin{gather}
\begin{aligned} \label{eq:disp_slow_highk}
   0 & =   \omega^2  \frac{2}{3} i \chi \eta \frac{\omega_B}{\omega_{\rm A}}  (2 k_\parallel^2 - k_\perp^2)  + 3i k_\perp^2 \omega_B  \Delta  \\ & + 2 \chi \eta \frac{\omega_B}{\omega_{\rm A}} k_\parallel \omega_{\rm ff} c_s  \Big( -k^2  (\bm{\hat{b}\cdot \hat{g}}) + (\bm{k \cdot \hat{g}})k_\parallel \Big)   \\ &  -
    \omega_{\rm A} {\rm v_A} k^2 k_\parallel  \frac{4}{3}i\eta \chi \frac{\omega_B}{\omega_{\rm A}}   .
\end{aligned}
\end{gather}

 \begin{figure}
  \centering
  \begin{minipage}[b]{\textwidth}
    \includegraphics[width=0.45\textwidth]{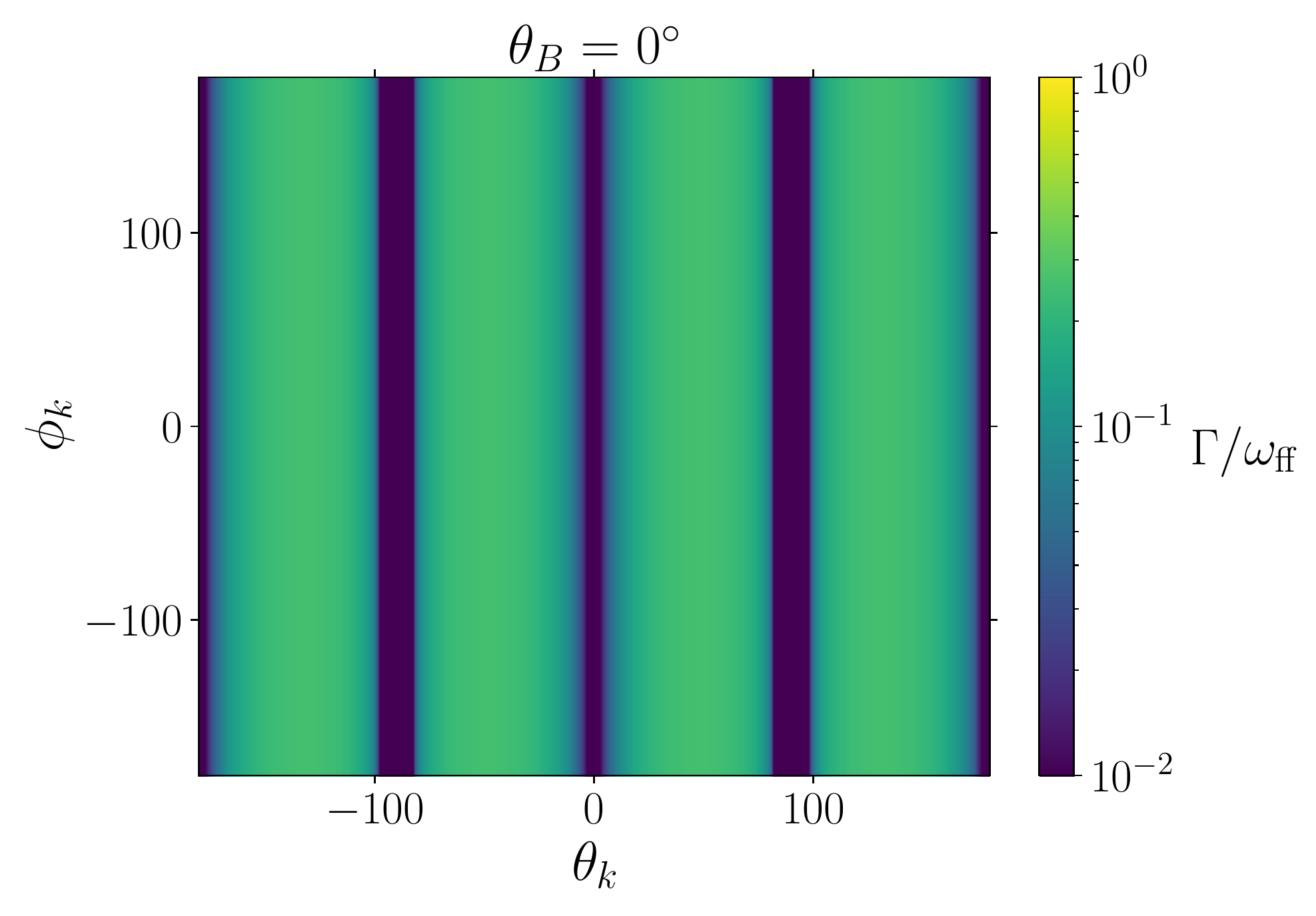}
  \end{minipage}
        \begin{minipage}[b]{\textwidth}
    \includegraphics[width=0.45\textwidth]{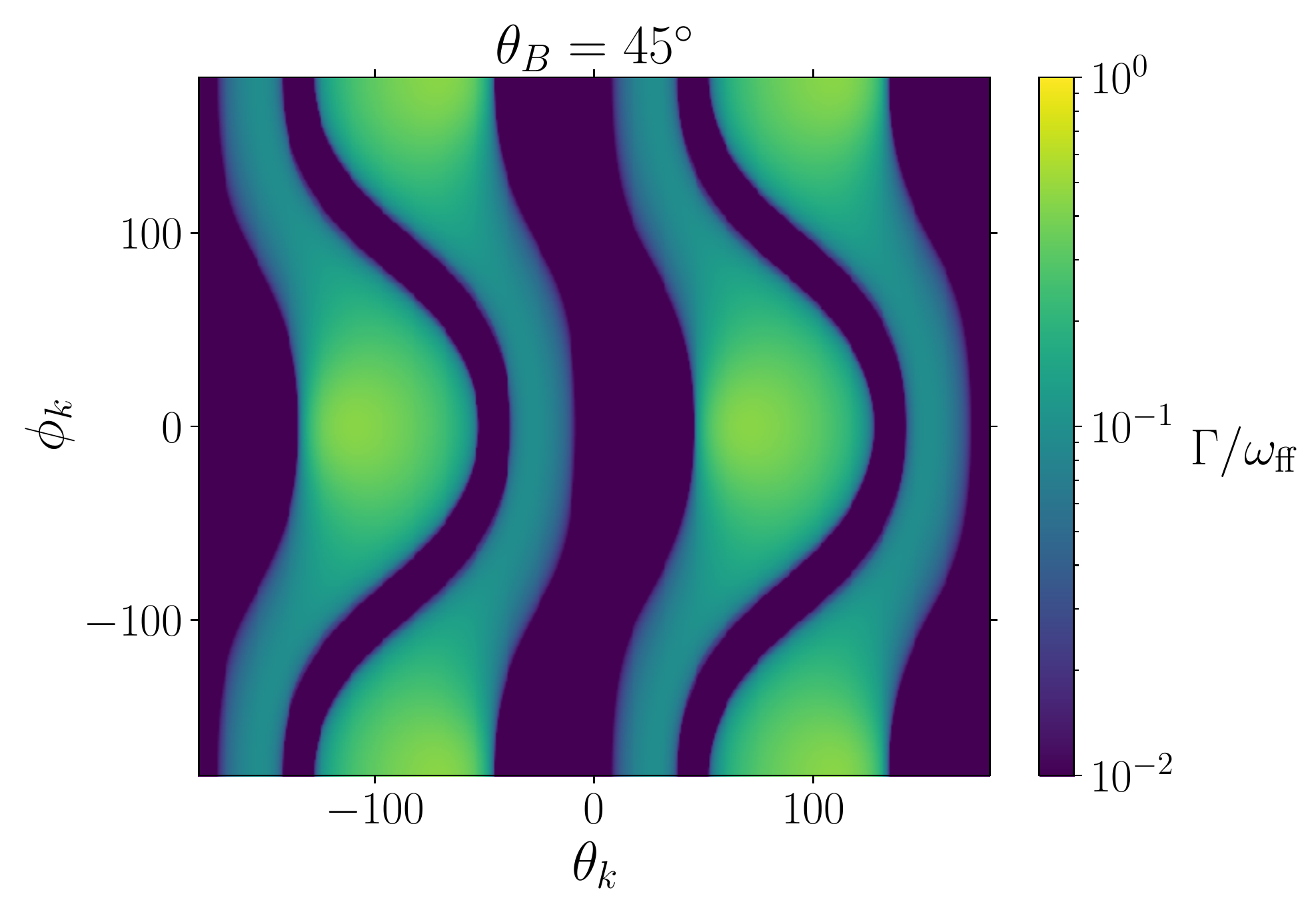}
  \end{minipage}
 \begin{minipage}[b]{\textwidth}
    \includegraphics[width=0.45\textwidth]{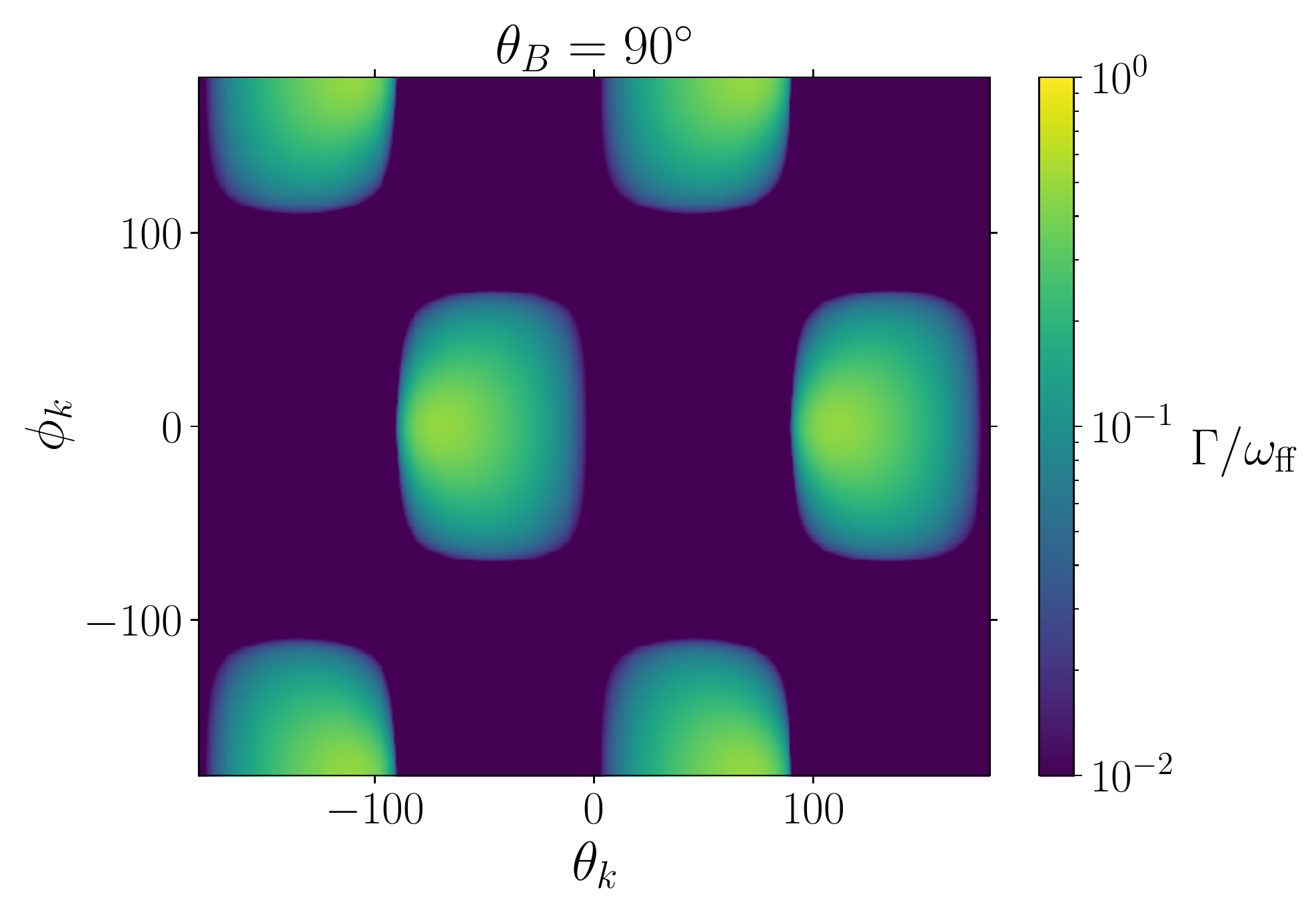}
  \end{minipage}
  \caption[CRBI growth rates as a function of propagation direction.]{CRBI growth rates as a function of propagation direction for $\eta = 0.1$, $\beta=100$, $H = 100 l_{\rm mfp}$ and fixed $kH=20$. The angles $\theta_k$ and $\phi_k$ are defined such that $k_x = k \sin \theta_k \cos \phi_k$, $k_y = k \sin \theta_k \sin \phi_k$ and $k_z = k \cos \theta_k$ (all the other figures in this paper use $\phi_k =0$). The three panels show growth rates for different orientations of the background magnetic field relative to gravity, which is in the $-\bm{\hat{z}}$ direction. The growth rates have a qualitatively different angular dependence for the different field geometries (see equations \ref{eq:gamma}, \ref{eq:gamma_ver} and \ref{eq:gamma_hor}). \label{fig:angle}}
\end{figure}

\subsection{Growth rates of short-wavelength modes} \label{sec:short}
We split the eigenmode frequency into real and imaginary parts, $\omega = \omega_{\rm R} + i \Gamma$, such that $\Gamma >0$ corresponds to exponential growth. We note that if magnetic tension and damping by pressure anisotropy in eq. \eqref{eq:disp_slow_highk} can be ignored (4th and 2nd term, respectively), we recover the growth rate that was derived in Section \ref{sec:gravity_simple}, i.e. $\Gamma \sim \sqrt{\omega_s \omega_{\rm ff}}$. From \eqref{eq:disp_slow_highk} we see that damping by anisotropic viscosity (pressure) can be ignored if the third term $\propto \omega_{\rm ff}$ is much larger than the damping term, or
\begin{equation} \label{eq:no_dp_condition}
    \beta \gg \eta^{-2} \Big(\frac{\omega_{s}}{\omega_{\rm ff}} \Big).
\end{equation}
At sufficiently high $\beta$ there is therefore a finite range of scales where the simple model from Section \ref{sec:gravity_simple} and eq. \eqref{eq:simple_LF} correctly predict the solution (see Figure \ref{fig:simple_picture}). 


We proceed by solving \eqref{eq:disp_slow_highk} in the limit $\eta \ll 1$, i.e. for small CR pressure fractions. To leading order, $\Delta =0$ is a solution, which, ignoring CR diffusion, implies
\begin{equation}
    \omega \approx \chi \omega_{\rm A}.
\end{equation}
This mode is the CR entropy mode. We stress again that the dependence on $\omega_{\rm A}$ in the CR entropy mode does not come from the perturbed magnetic tension, but from CR streaming along field lines at the Alfv\'en speed, which also has characteristic frequency $\omega_{\rm A}$. The growth rate of the mode can be found at first order in $\eta$ and is approximately given by,
\begin{equation} \label{eq:gamma}
    \Gamma \approx \chi \frac{\sqrt{2}}{3} \eta \beta^{1/2} \omega_{\rm ff} \frac{-k^2  \bm{\hat{b} \cdot \hat{g}} + \bm{k \cdot \hat{g}} k_\parallel}{k_\perp^2},
\end{equation}
where $\chi = \pm 1$ is the parameter that ensures that CR stream down their pressure gradient (eq. \ref{eq:va_mod}).
The growth rate increases with increasing $\beta$, as at higher $\beta$ the pressure anisotropy is better minimised (eq. \ref{eq:pbal}) and anisotropic viscous damping is reduced. There is no unstable growth if gravity is normal to the $\bm{k}-\bm{B}$ plane. In the case of a magnetic field that is antiparallel to $\bm{g}$,
\begin{equation} \label{eq:gamma_ver}
    \Gamma \sim \chi \eta \beta^{1/2} \omega_{\rm ff} ,\quad \bm{B} = B \bm{\hat{z}}, \ \bm{g} = -g \bm{\hat{z}}.
\end{equation}
For a horizontal magnetic field along $x$,
\begin{equation} \label{eq:gamma_hor}
    \Gamma \sim -  \chi \eta \beta^{1/2} \omega_{\rm ff} \frac{k_z k_x}{k_z^2 + k_y^2} ,\quad \bm{B} = B \bm{\hat{x}}, \ \bm{g} = -g \bm{\hat{z}}.
\end{equation}
While for $\bm{B} = B \bm{\hat{z}}$ and $\chi>0$ all modes are unstable (except when $\bm{k}$ is approximately parallel to or perpendicular to $\bm{B}$), for $\bm{B} = B \bm{\hat{x}}$ growth occurs if $\chi k_x  k_z  < 0$. Unless $\bm{B} \parallel \bm{g}$ and $\chi \bm{B \cdot g} > 0$ ($dp_c / dz > 0$, which is unlikely), there exists a region in $k$-space where there is wave growth. We note that for horizontal magnetic fields growth rates can be higher than for vertical magnetic fields because of the extra factor that depends on the direction of propagation, although growth rates with $k_x \gg k_y,k_z$ do not diverge because the ordering used to derive \eqref{eq:gamma}--\eqref{eq:gamma_hor} breaks down. 

The above solutions are for pure CR streaming and also do not include the impact of background gradients on the growth rates. By neglecting CR diffusion we have assumed that CRs are perfectly coupled to the self-excited Alfv\'en waves and so stream at the Alfv\'en speed. We include CR diffusion in our analysis to relax the assumption of pure Alfv\'enic streaming, which certainly breaks down on small scales. From eq. \eqref{eq:disp_slow}, or from the fact that in the presence of diffusion the CRs' natural frequency is $\omega \approx \omega_{\rm A} - i \omega_d$ (see eq. \ref{eq:pc_en2}), we can estimate that CR diffusion suppresses the instability when
\begin{equation}
    \omega_d \gtrsim \eta \beta^{1/2} \omega_{\rm ff}
\end{equation}
or equivalently

\begin{equation} \label{eq:diffusion_k}
k l_{\rm mfp} \gtrsim  \Big(\Phi^{-1} \frac{l_{\rm mfp}}{H} \eta \beta^{1/2} \Big)^{1/2},
\end{equation}
where we used the parameter $\Phi$ defined in eq. \eqref{eq:phi}.

\eqref{eq:pc_en2} also allows us to estimate how background gradients affect the growth rate in \eqref{eq:gamma}. The background CR pressure gradient modifies the mode's imaginary part by ${\rm v_A}_{,z} d \ln p_c / d z \sim {\rm v_A}_{,z}/H \sim \omega_{\rm ff}/ \beta^{1/2}$, which encapsulates that as the perturbation propagates down the CR pressure gradient, $\delta p_c / p_c(z)$ remains constant in collisional MHD without CR diffusion. The background therefore significantly modifies the growth rate if,
\begin{equation}
   \frac{ \omega_{\rm ff} } {\beta^{1/2}} \gtrsim \eta \beta^{1/2} \omega_{\rm ff} \ \ \Longrightarrow \ \ \eta \lesssim \beta^{-1}.
\end{equation}
We stress again that for $\eta \lesssim \beta^{-1}$ the mode is not damped in the usual sense, because it maintains approximately constant $\delta p_c / p_c(z)$ as it propagates down the CR pressure gradient. 

 \begin{figure}
  \centering
    \includegraphics[width=0.45\textwidth]{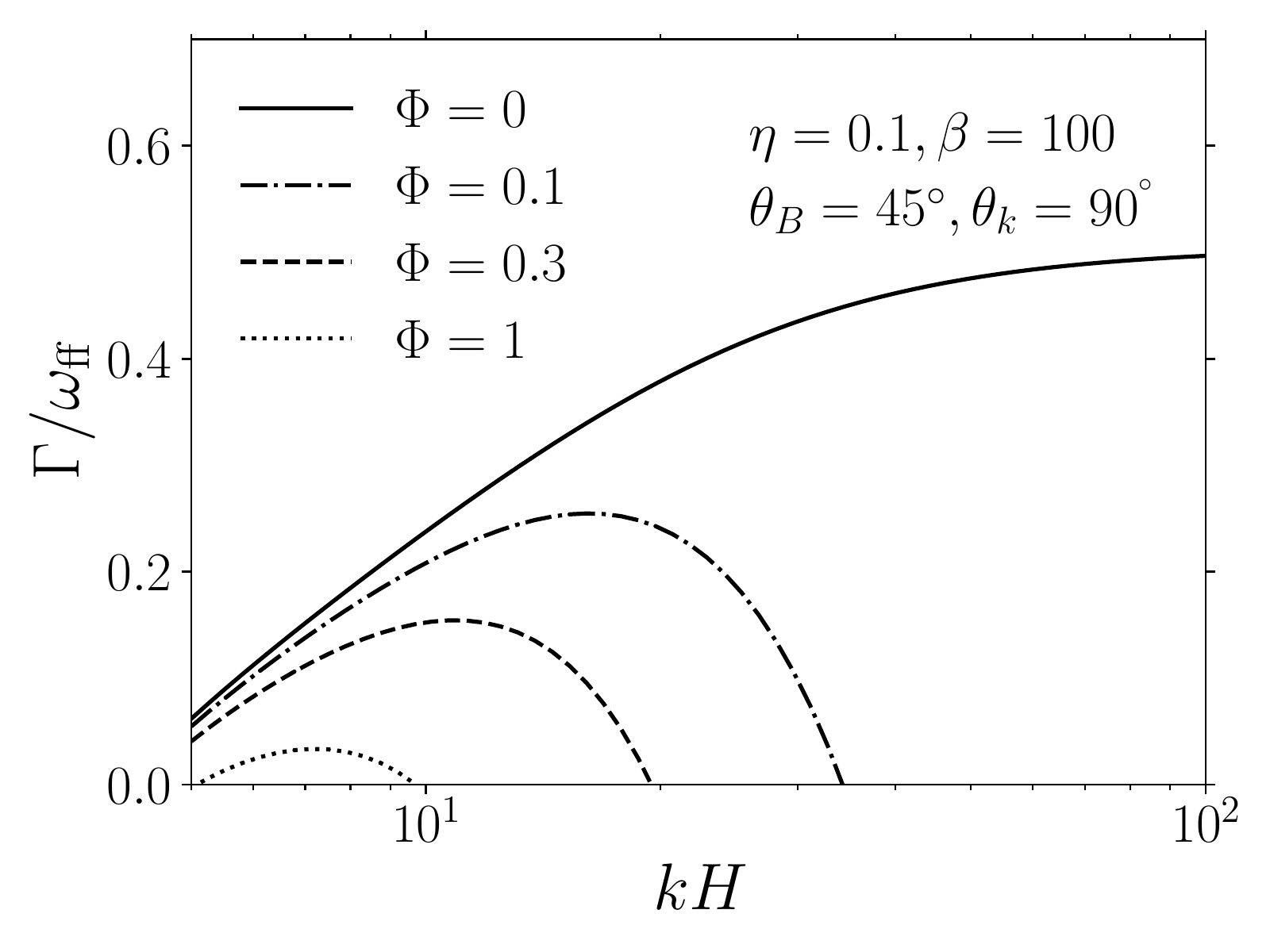}
  \caption[Impact of CR diffusion on CRBI growth rates.]{CRBI growth rates as a function of wavenumber and the CR diffusion coefficient (quantified using the parameter $\Phi$ defined in eq. \ref{eq:phi}). We assume $\eta = 0.1$, $\beta = 100$, $H = 100 l_{\rm mfp}$ and the isothermal background described in Section \ref{sec:icm}. CR diffusion suppresses growth at high $k$ (eq. \ref{eq:diffusion_k}) and can completely shut off the instability if sufficiently large (eq. \ref{eq:phi_cond}). \label{fig:diff}}
\end{figure}

 \section{CR Buoyancy Instability in an Isothermal Atmosphere} \label{sec:icm}
We now complement the analytics of Section \ref{sec:grav} with numerical solutions of equations \eqref{eq:drho}-\eqref{eq:dpc} including background gradients. In this section, we consider an isothermal atmosphere in order to isolate the CRBI from the HBI/MTI, which require background temperature gradients.  As our fiducial set of parameters, we use $\eta=0.1$, $\beta = 100$ and $H = 100 l_{\rm mfp}$. The background is in the hydrostatic equilibrium described by \eqref{eq:back_general_HE} and \eqref{eq:back_general_pc}. We assume that the background CR heating is balanced by an unspecified cooling function, which we do not perturb in our linear analysis, except in Section \ref{sec:cool} (see \citealt{kq2020} for a discussion of thermal instability with streaming CRs). Unless specified otherwise, we will consider wavevectors $\bm{k}$ in the $\bm{B}-{\bm g}$ plane, i.e. $\phi_k = 0$ (see Section \ref{sec:lin_eq}), motivated by the fact that the instability is not present if $\bm{g}$ is perpendicular to the  $\bm{B}-{\bm k}$ plane. We show how growth rates depend on $\phi_k$ in Figure \ref{fig:angle}.

The physics of the instability described in Section \ref{sec:prelims} becomes apparent by plotting the properties of the CRBI. We show the mode properties for $\eta=0.1$ and $\beta=100$ in Figure \ref{fig:mode}. We plot the oscillation frequency in panel a), the growth rate in panel b),  $k_\parallel \vrm_\parallel / k_\perp  \vrm_\perp$, which quantifies the compressibility of the mode ($=-1$ if incompressible), in panel c), and $[\delta p_c + \delta (B^2/8\pi)] / \delta p_g$, which quantifies the degree of pressure balance ($=-1$ if pressure balanced, i.e. $\delta p_c + \delta (B^2/8\pi) + \delta p_g = 0$), in panel d). The blue line shows the unstable CR entropy mode. For completeness, we also plot the MHD slow modes and the gas-entropy mode. Panel b) shows that all modes except the CR entropy mode are strongly damped by low-collisionality physics (viscosity and conduction).  At small $k$,  the CR entropy mode  is approximately incompressible ($k_\parallel {\rm v}_\parallel \approx - k_\perp {\rm v}_\perp$) and the growth rates are $ \ll \omega_{\rm ff}$.  At high $k$, the mode becomes compressible due to CR streaming,  approaches $k_\perp \vrm_\perp = 2 k_\parallel \vrm_\parallel$, and the growth rate reaches the plateau given by eq. \eqref{eq:gamma}. We note that $k_\perp {\rm v}_\perp$ does not quite reach  $2 k_\parallel {\rm v}_\parallel$ because we limit the x-axis to $k l_{\rm mfp} < 1$, where the Braginskii MHD model is valid. The oscillation frequency $\approx \omega_A$ is set by the characteristic frequency of CR streaming. 

In Figure \ref{fig:growth_vs_eta} we show growth rates of the CRBI for the fiducial parameters and different ratios of CR pressure to gas pressure, $\eta$. The instability exists even for small CR pressures, but with reduced growth rates. In the analytics in Section \ref{sec:grav} without background gradients, the instability exists for arbitrarily small CR pressures. Here we find that with background gradients, within our local WKB framework, the instability exists for any $\eta \gtrsim \beta^{-1}$ (ignoring CR diffusion), which is consistent with the discussion at the end of Section \ref{sec:short}.

We show growth rates as a function of propagation direction at fixed $kH = 20$ in Figure  \ref{fig:angle} for different orientations of the background magnetic field. The growth rates in the three panels have a qualitatively different angular dependence, consistent with equations \eqref{eq:gamma_ver} and \eqref{eq:gamma_hor}.  For horizontal magnetic fields growth rates are larger than for vertical magnetic fields, but growth occurs in a smaller region of $k$-space. 

In Section \ref{sec:short} we noted that significant CR diffusion suppresses the CRBI at short wavelengths. We now show this explicitly in Figure \ref{fig:diff} for different values of $\Phi$, which quantifies the strength of CR diffusion (eq. \ref{eq:phi}).\footnote{We assume that CR diffusion does not affect the background equilibrium, i.e. we assume an approximately linear CR pressure profile,  $ \kappa d^2 p_c /d z^2 \approx 0$.}  As predicted by \eqref{eq:diffusion_k}, diffusion suppresses the instability at high $k$. 

\begin{figure}
  \centering
  \begin{minipage}[b]{\textwidth}
    \includegraphics[width=0.45\textwidth]{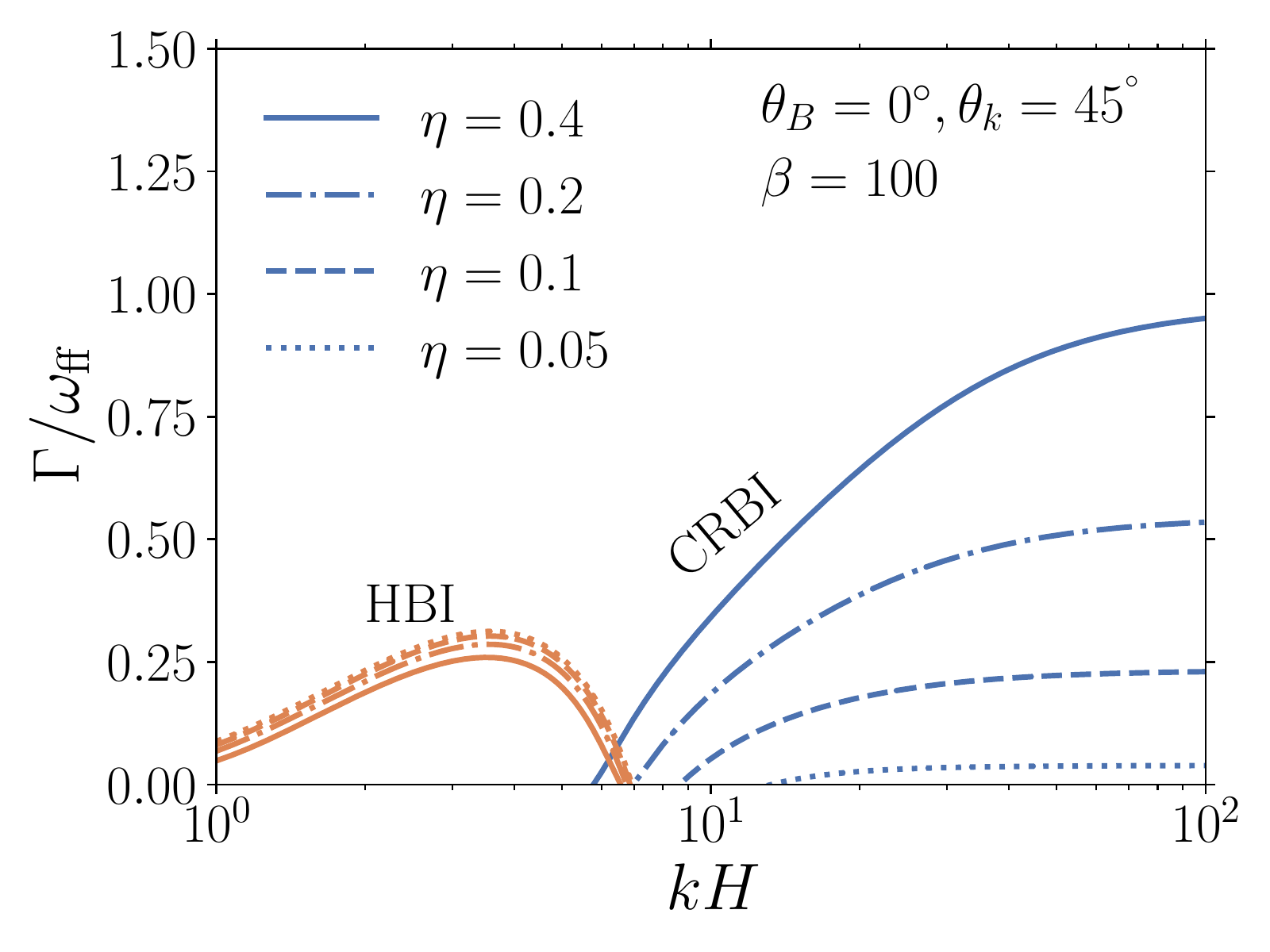}
  \end{minipage}
        \begin{minipage}[b]{\textwidth}
    \includegraphics[width=0.45\textwidth]{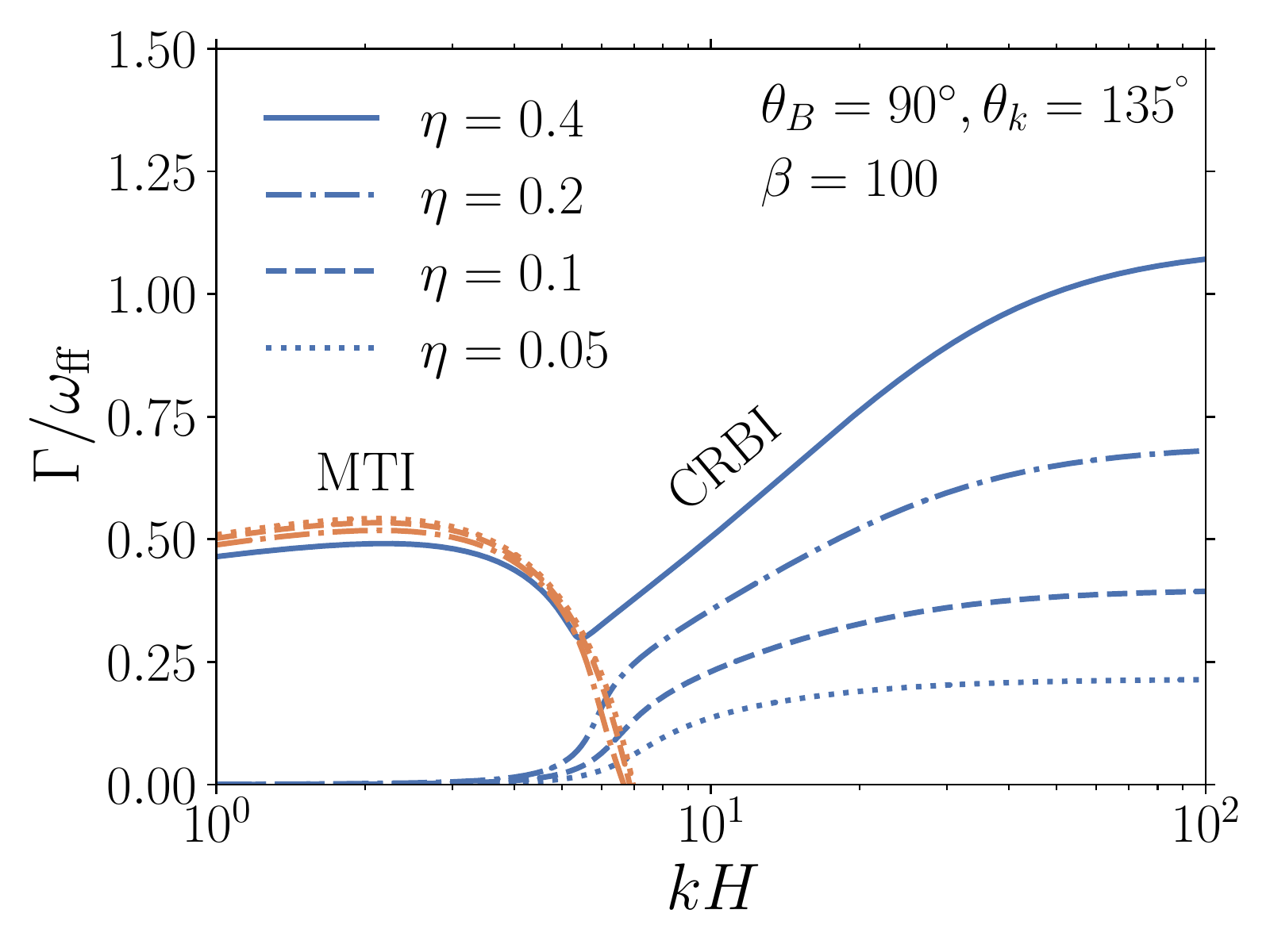}
  \end{minipage}
  \caption[Comparison of the CRBI to the HBI and MTI.]{CRBI versus buoyancy instabilities driven by background stratification and anisotropic heat conduction, the HBI (top) and MTI (bottom). We use $\beta = 100$, $H=100l_{\rm mfp}$ and different $\eta$. Long-wavelength modes are destabilised by thermal conduction, as in the MTI/HBI. Short-wavelength modes are stable to the HBI/MTI due to the stabilising effect of magnetic tension. Short-wavelength modes are, however, destabilised by the compressibility induced by CR streaming. The CRBI  therefore operates alongside the long-wavelength HBI and MTI. For $\eta=0.4$, the CRBI  and the MTI are destabilising the same mode (also true for the HBI at higher $\beta$).   \label{fig:hbi_mti}}
\end{figure}

Given that the CRBI typically becomes important on scales $kH \gg 1$ and that at longer wavelengths buoyancy instabilities such as the HBI and MTI are generally more important (see Section \ref{sec:hbi_mti} and Figure \ref{fig:hbi_mti}), we can rephrase \eqref{eq:diffusion_k} in terms of a rough overall criterion for the suppression of the CRBI by CR diffusion (however, we note that for large thermal mean free paths, as in cluster outskirts, the CR-driven instability is important on scales $kH\sim 1$, see Figure \ref{fig:lf}). Setting $k_{\rm min}H$ as the largest scale on which the CRBI operates and using \eqref{eq:diffusion_k}, we find that CR diffusion suppresses the instability if,
\begin{equation} \label{eq:phi_cond}
    \Phi \gtrsim 0.01 \eta \beta^{1/2}  \frac{H }{l_{\rm mfp}} \Big(\frac{k_{\rm min} H}{10} \Big)^{-2}.
\end{equation}
For $\eta = 0.1$, $\beta = 100$, $k_{\rm min}H=10$ and $H= 100 l_{\rm mfp}$, CR diffusion suppresses the instability for $\Phi \gtrsim 1$, roughly consistent with Figure \ref{fig:diff}.

\section{Discussion} \label{sec:discussion}
\subsection{CRBI versus HBI/MTI} \label{sec:hbi_mti}
We now consider the relationship between the CRBI and previously identified buoyancy instabilities driven by background temperature gradients and heat fluxes, i.e. the MTI (\citealt{mti_balbus}) and the HBI (\citealt{hbi}). How CRs may affect these buoyancy instabilities has been considered in previous work (e.g., \citealt{cd06}, \citealt{dc09}, \citealt{sharma_cr_buoyancy}). However, they did not use streaming CR transport and so the CRBI was not included in their calculation. 

Instead of an isothermal atmosphere as in Section \ref{sec:icm}, we here consider a background temperature that increases with height, as is the case in cluster cores: 
 \begin{equation} \label{eq:hbi_back}
     \frac{d}{dz} (p_g + p_c) = - \rho g, \quad p_c \propto \rho^{2/3}, \quad \frac{d \ln T}{dz} =  H^{-1} ,
 \end{equation}
where $H = c_s^2 /g =  c_s / \omega_{\rm ff}$. This equilibrium with vertical magnetic field and $T$ increasing with height is unstable to the HBI at high $\beta$ (driven by the background anisotropic heat flux).  We also consider a background with $dT/dz < 0$ and a horizontal magnetic field, $\bm{B} = B \bm{\hat{x}}$, which is unstable  to the MTI. To study the MTI we consider the following background,
\begin{equation}\label{eq:mti_back}
     \frac{d p_g}{dz} = - \rho g, \quad \rho = {\rm const}, \quad \frac{d \ln T}{dz} = -  H^{-1} =  - \frac{\omega_{\rm ff}}{ c_s}.
 \end{equation}
For the MTI, we assume that $|\bm{\hat{b} \cdot \nabla } p_c |/p_c  \ll H^{-1}$, so that the cosmic rays are coupled but their background gradient is sufficiently small to be ignored (which is consistent with our choice of $\rho \approx {\rm const}$). We stress that this choice is made for the sake of simplicity and is not necessarily representative of cluster conditions. 

We show growth rates for the backgrounds described by equations \eqref{eq:hbi_back} and \eqref{eq:mti_back} in Figure \ref{fig:hbi_mti}, for $\beta = 100$, $H/l_{\rm mfp} = 100$ and different values of $\eta$. At small $k$, growth rates are dominated by the HBI/MTI. At high $k$, the HBI is partially stabilized by anisotropic viscosity (\citealt{kunz_mti}), and the MTI and HBI are completely suppressed  by magnetic tension. The HBI is also partially suppressed by the CR pressure gradient at long wavelengths (for large $\eta$), where CRs are approximately adiabatic, $\omega_{\rm A} < \omega_{\rm ff}$ (\citealt{sharma_cr_buoyancy}).  Short-wavelength modes are  destabilised by compressibility induced by CR streaming. We note that in the top panel there is a range of wavelengths where our calculation does not predict unstable growth at small $\eta$, which is not the case in the bottom panel. This is due to the effect of the background CR pressure gradient on the growth rate, explained in Section \ref{sec:short}, which is not present in the equilibrium used to study the MTI (eq. \ref{eq:mti_back}). Finally, we note that for $\eta=0.4$  the CRBI and the MTI are in fact driving the same mode (this is also true for the HBI for $\beta$ slightly larger than used in Figure \ref{fig:hbi_mti}).

The CRBI considered in this work therefore operates alongside standard buoyancy instabilities driven by background gradients, such as the HBI/MTI. Both types of instabilities are driven by gravity acting on density fluctuations. At long wavelengths, for which the HBI/MTI operate, the density fluctuations that introduce unstable buoyancy are due to a combination of background density stratification, rapid heat conduction and pressure balance. At short wavelengths, the unstable density fluctuations are due to CR streaming and pressure balance, independent of the background stratification.

The transition from heat-flux-driven growth to CR-driven growth in Figure \ref{fig:hbi_mti} occurs around $kH \sim 5$. The exact value is sensitive to our choice of parameters, such as $\eta$ or $\beta$. It also depends strongly on the thermal mean free path, more specifically the ratio $H/l_{\rm mfp}$, which sets the range of $k$ for which CR streaming drives the mode away from incompressibility. In particular, for $H \sim l_{\rm mfp}$ the CRBI can have faster growth rates than the HBI/MTI at long wavelengths (see Figure \ref{fig:lf}) for plausible parameters. We discuss the dependence of the CRBI on the value of the thermal particle mean free path in more detail in Section \ref{sec:collisionless}.

\subsection{Impact of cooling} \label{sec:cool}

 \begin{figure}
  \centering
    \includegraphics[width=0.45\textwidth]{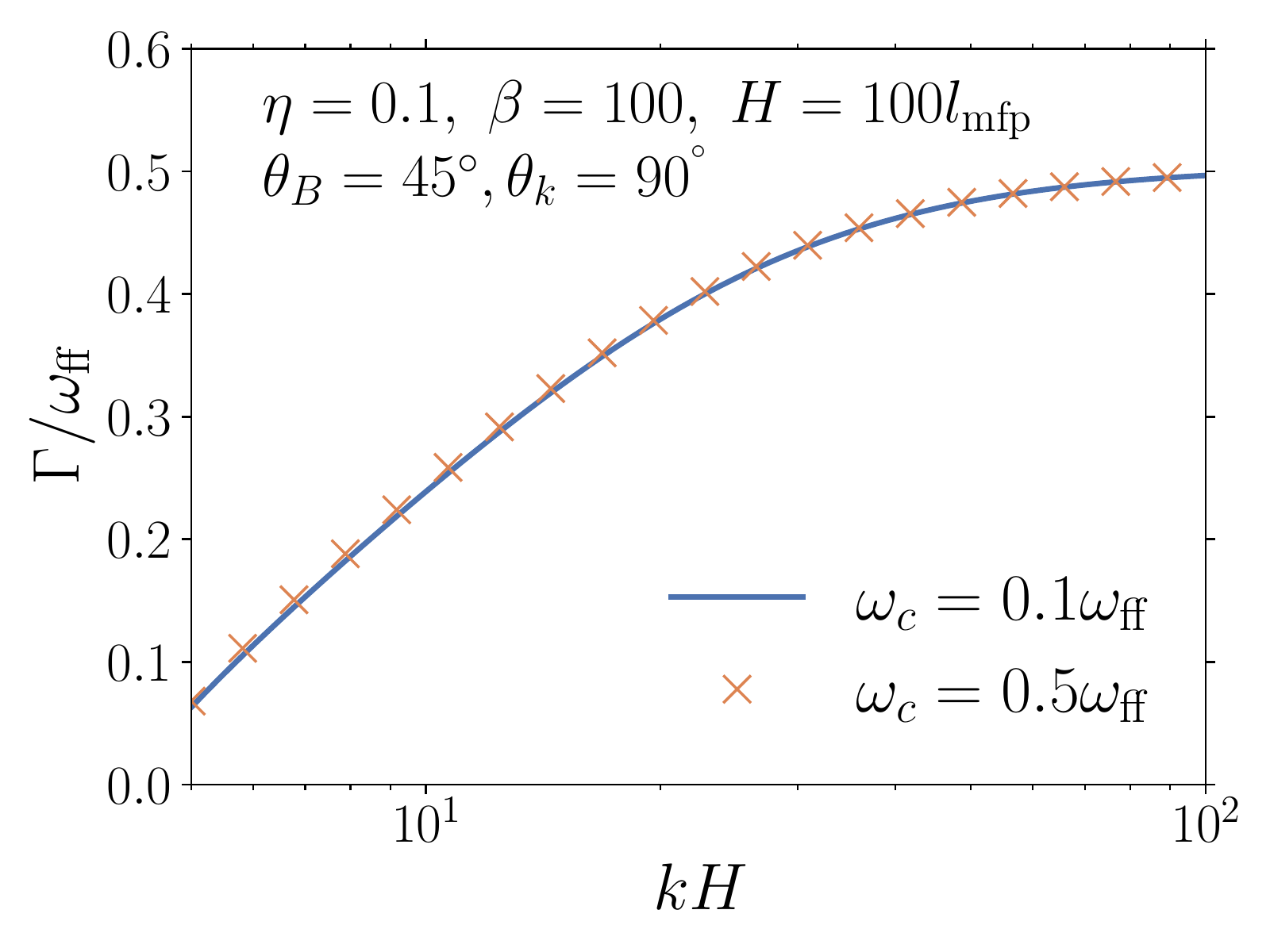}
  \caption[Growth rates of the CRBI for different cooling rates.]{ Growth rates of the CRBI are not significantly affected by cooling even when the cooling rate $\omega_c$ is comparable to the growth rate. This is because the mode is isothermal due to rapid conduction (the wavelengths shown are below the Field length where thermal instability is suppressed by conduction; \citealt{field65}). For this plot, we assume thermal Bremsstrahlung to be the dominant radiative cooling process.\label{fig:cool}}
\end{figure}

We have ignored cooling throughout this work. Given that the unstable short-wavelength CR entropy modes have significant density fluctuations due to CR streaming, cooling could in principle have an impact on the instability. However, because the unstable wavelengths are characterised by thermal-conduction times that are much shorter than the cooling time, the dominant response of the gas is simply that it is isothermal, even in the presence of cooling and large CR-driven density fluctuations. The perturbed cooling therefore has no significant effect on the CRBI even when the cooling rate $\omega_c$ is comparable to the growth rate, as we show in Figure \ref{fig:cool}.

\subsection{Dilute cluster outskirts and the collisionless regime} \label{sec:collisionless}
In Figures \ref{fig:mode}--\ref{fig:cool} we used a fixed $H/l_{\rm mfp} = 100$. While $H / l_{\rm mfp} \gg 1$ is representative of the conditions in the inner regions of galaxy clusters, $H/l_{\rm mfp}$ is likely smaller in the outskirts, where the ICM plasma density is significantly reduced. A larger mean free path implies that CR-streaming-induced compressibility effects become important on larger scales; the growth rates of long-wavelength modes will thus be enhanced relative to the results from Figures \ref{fig:mode}--\ref{fig:cool}.

 \begin{figure}
  \centering
    \includegraphics[width=0.45\textwidth]{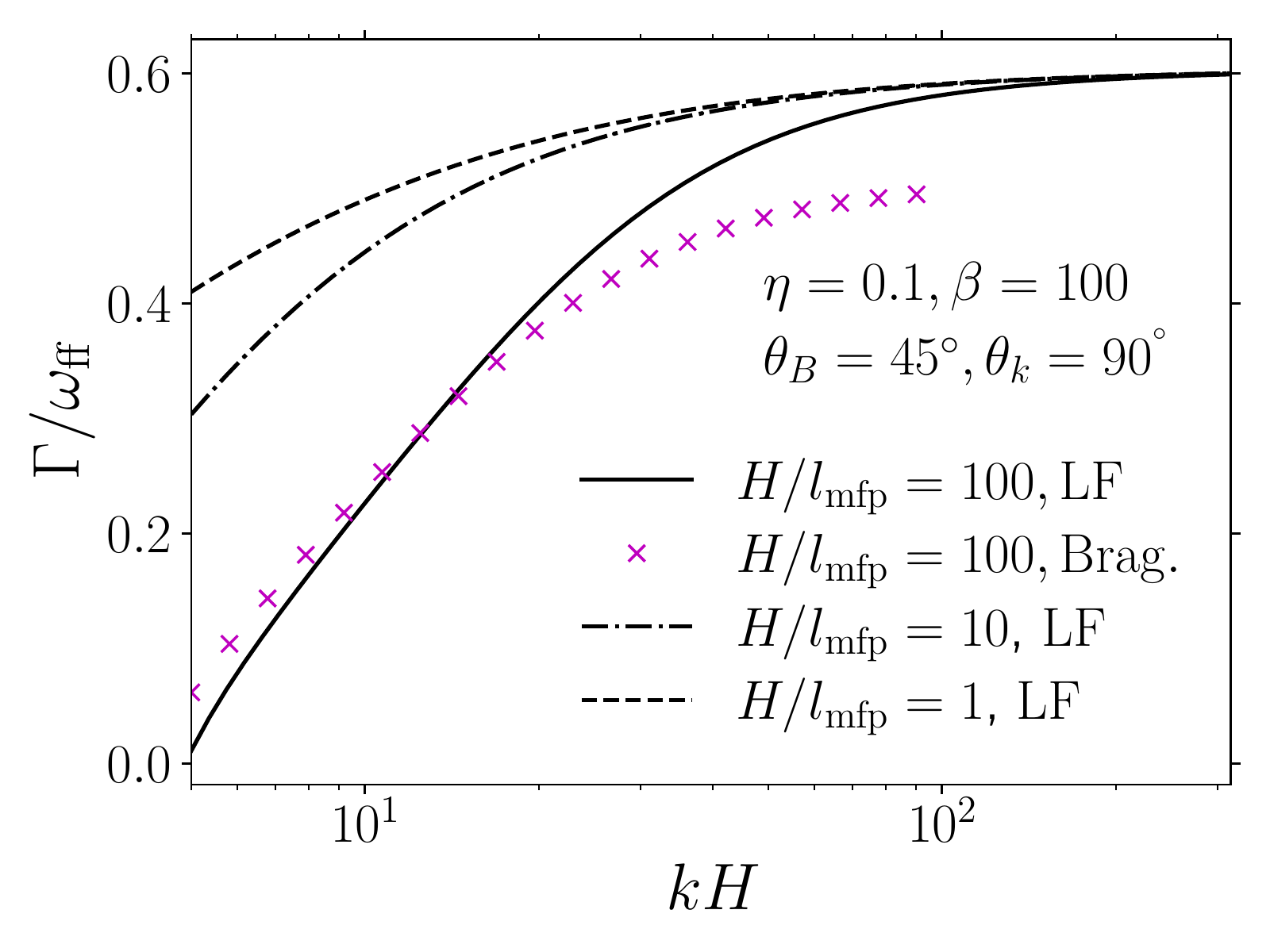}
  \caption[Growth rates of the CRBI for different collisionalities.]{Growth rates of the CRBI, calculated using the Landau-fluid model (black lines), for $\eta=0.1$, $\beta =100$ and different $H/l_{\rm mfp}$. The Landau-fluid model allows us to compute approximate growth rates for both the collisional and collisionless regimes. For larger $l_{\rm mfp}$, as is likely the case in dilute cluster outskirts, the growth rates of long-wavelength modes are enhanced, as CR streaming leads to significant density fluctuations on large scales. There is good agreement between Landau-fluid and Braginskii MHD (magenta crosses) predictions in the collisional regime, as expected. \label{fig:lf}}
\end{figure}

Considering a larger mean free path runs into the issue that the range of scales for which both the Braginskii MHD model and the WKB approximation  are valid ($k l_{\rm mfp} \ll 1$ and $k H \gg 1$, respectively) becomes very limited. To alleviate this issue, we here consider a different model for the low-collisionality thermal plasma. We use the kinetic MHD equations (\citealt{cgl}) with a  ``Landau-fluid" prescription for the heat fluxes, i.e. the heat fluxes are constrained by the requirement that the fluid equations approximately match the linear response of the kinetic thermal plasma (\citealt{snyder97}). We use the heat fluxes from \cite{snyder97} that depend on the collision rate, allowing for a smooth transition between the weakly-collisional (Braginskii MHD) and collisionless regimes.  The equations of the Landau-fluid model are provided in Appendix \ref{app:lf}.  For simplicity, we ignore electron physics and only consider the thermal ions, which dominate the pressure anisotropy because the ion collision rate is much smaller than the electron collision rate. Ignoring the effect of the electron heat flux on the ions is generally not rigorous (in our Braginskii calculation the heat flux was due to electrons, hence ${\rm Pr =0.02}$). However, as discussed in Section \ref{sec:validity_thermal_gas}, the electron-ion thermal equilibration rate is  $\tau_{\rm eq}^{-1} \sim (m_e / m_i)^{1/2} \nu_{\rm ii}\sim \nu_{\rm ii}/40$, i.e. ions and electrons are thermally decoupled for $\omega \sim \omega_{\rm A}$ modes if $k l_{\rm mfp} \gtrsim \beta^{1/2} / 40$. For $\beta \sim 100$ and $k l_{\rm mfp} \gtrsim 1$ electrons and ions are then approximately thermally decoupled. It is therefore reasonable to neglect the electron heat flux in the collisionless regime, which is of primary interest in this section.

We show growth rates of the CRBI, calculated using the Landau-fluid model (black lines), in Figure \ref{fig:lf} for $\eta=0.1$, $\beta =100$ and different $H/l_{\rm mfp}$. As expected, for larger $l_{\rm mfp}$ the growth rates of long-wavelength modes are enhanced, as CR streaming leads to larger density fluctuations on large scales. The asymptotic growth rate at high $k$ is independent of the mean free path (absent CR diffusion). We also compare the Landau-fluid results to the Braginskii MHD calculation for the fiducial case $H / l_{\rm mfp}=100$ (solid line and magenta crosses in Figure \ref{fig:lf}). There is good agreement between the two models in the collisional regime, which shows that instability growth rates are not very sensitive to the thermal Prandtl number, as the Braginskii MHD model has ${\rm Pr}=0.02$ while the Landau-fluid model has ${\rm Pr} \sim 1$ (this is consistent with our finding that ${\rm Pr}=0.02$ and ${\rm Pr}\sim 1$ yield similar results in Braginskii MHD). The asymptotic high-$k$ growth rates in the two models are also remarkably similar. This is because the qualitative physical picture of the instability does not change between the collisional and collisionless regimes (although the exact relationship satisfied by $k_\parallel {\rm v}_\parallel$ and $k_\perp {\rm v}_{\perp}$ at high $k$ is different in the two models, and in the Landau-fluid model $k_\parallel {\rm v_\parallel} \gg k_\perp {\rm v}_{\perp}$).

\subsection{Diffusive correction to CR streaming} \label{sec:diffusion}
Figure \ref{fig:diff} shows that significant CR diffusion can suppress the CRBI. The magnitude of the diffusive correction to Alfv\'enic streaming is therefore critical. The diffusive correction depends on the damping of the Alfv\'en waves excited by the CR streaming instability. In the ICM, the dominant damping mechanisms are nonlinear Landau damping $\sim k {\rm v_{th}} (\delta B / B)^2$ (\citealt{lee_volk_1973}; \citealt{kulsrud_book}) and linear Landau damping of Alfv\'en waves in a turbulent  background (\citealt{wiener_hib}). For turbulence injected on $\sim 10$ kpc (a common scale for the radio bubbles) with perturbations comparable to the Alfv\'en speed, the linear-Landau damping rate of $k \sim r_L^{-1}$ Alfv\'en waves excited by GeV CRs (where $r_L$ is the GeV CR gyroradius) is
\begin{equation}
    \Gamma_{\rm L} \sim \frac{0.4 {\rm v}_{\rm th}}{(r_L L_{\rm turb})^{1/2}} \sim 10^{-10} \ {\rm s^{-1}} \frac{{\rm v}_{\rm th}}{10^8 {\rm cm \ s^{-1}}} \Big(\frac{L_{\rm turb}}{10 \ {\rm kpc}}\Big)^{-1/2} \Big(\frac{B}{1 \ {\rm \mu G}} \Big)^{1/2}.
\end{equation}
We compute the correction to Alfv\'enic streaming for a combination of linear and nonlinear damping mechanisms in Appendix \ref{sec:diff_appendix}. The resulting diffusion coefficient is a function of the background CR pressure gradient.  We split the total diffusion coefficient $\kappa$ into two components, $\kappa (\nabla p_c)  = \kappa_{\rm diff} (\nabla p_c)$ + $\kappa_{\rm st} (\nabla p_c)$, where $\kappa_{\rm st}$ is the part of the diffusion coefficient that scales as $\kappa_{\rm st} \propto (\nabla p_c)^{-1}$ and therefore does not result in real diffusive behaviour (as needed to suppress the CRBI). For purely linear damping mechanisms, $\kappa = \kappa_{\rm st}$ (\citealt{skilling71}). Diffusive behaviour in the form of a finite $\kappa_{\rm diff}$ comes from non-zero nonlinear damping.   

\begin{figure}
  \centering
    \includegraphics[width=0.45\textwidth]{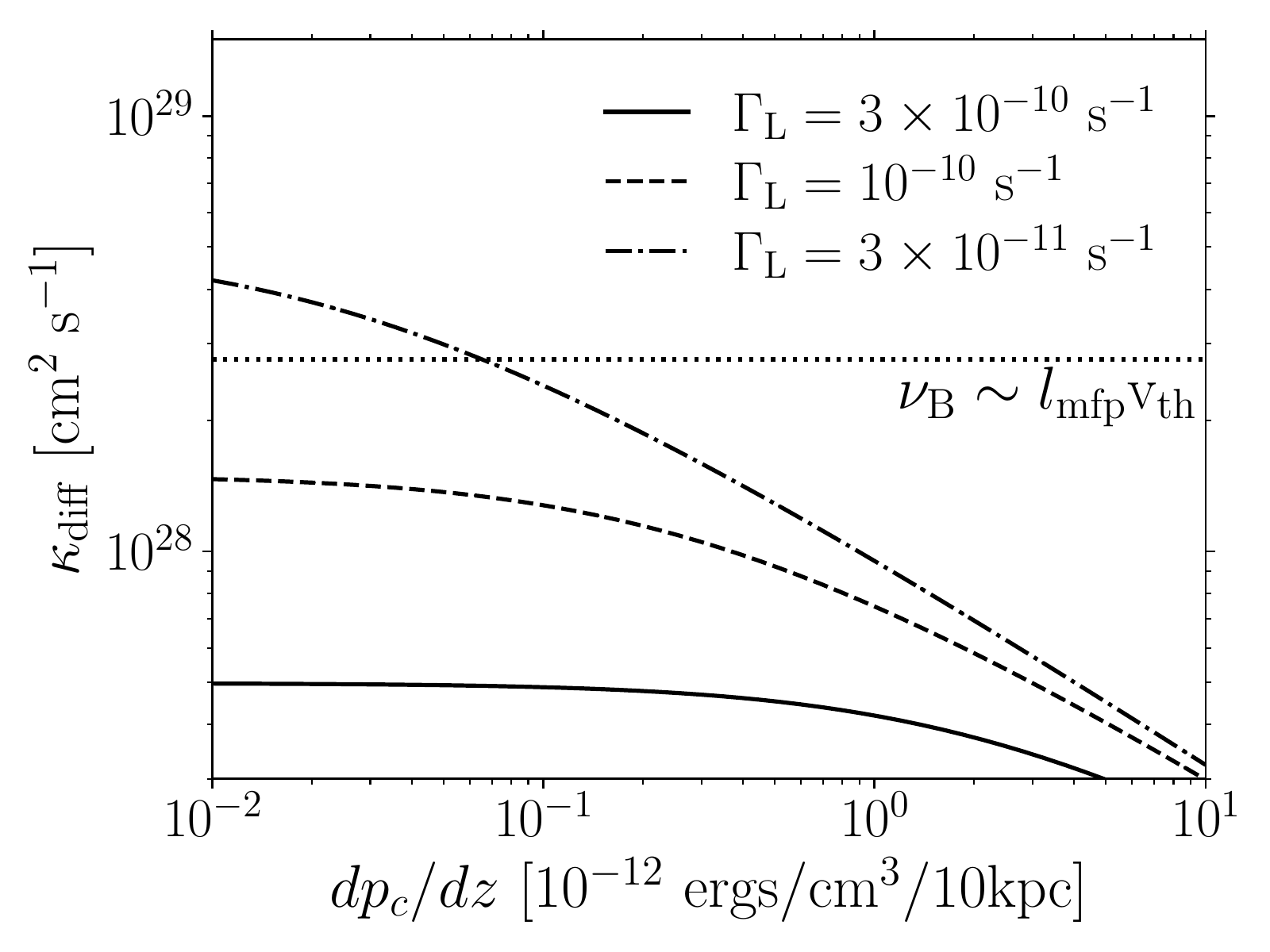}
  \caption[Diffusive component of the CR transport correction in self-confinement theory for different linear damping rates.]{ The diffusive correction to Alfv\'enic streaming calculated from eq. \eqref{eq:quad_balance} as a function of the CR pressure gradient, for different magnitudes of the linear damping of Alfv\'en waves excited by the CR streaming instability. The horizontal dotted line is the anisotropic viscosity of the thermal gas for $l_{\rm mfp} \sim 0.2 \ {\rm kpc}$ and $T=3 \times 10^7$K. Thus, $\Phi < 1$ (see eq. \ref{eq:phi}) is plausible in cluster cores and the CRBI is likely only partially suppressed by CR diffusion (Figure \ref{fig:diff}).   \label{fig:kappa_different_gamma} }
\end{figure}

$\kappa_{\rm diff}$ is plotted for different linear damping strengths in Figure \ref{fig:kappa_different_gamma} as a function of the CR pressure gradient, normalized using $p_c = 10^{-12} \  {\rm ergs / cm^3}$ and a scale height $H_c = 10 \ {\rm kpc}$. In addition to linear damping, the waves excited by the streaming instability are damped by nonlinear Landau damping. We plot $\kappa_{\rm diff}$ rather than the total $\kappa =  \kappa_{\rm diff} + \kappa_{\rm st} $ because $\kappa_{\rm diff}$ is the component that acts as a diffusion term (see Appendix \ref{sec:diff_appendix}). The non-diffusive correction to Alfv\'enic streaming $\kappa_{\rm st}$ likely does not suppress the CRBI and is a small correction to Alfv\'enic streaming for the GeV CRs in a steady state (\citealt{kq2022}). 

The horizontal dotted line in Figure \ref{fig:kappa_different_gamma} shows the Braginskii viscosity of the thermal plasma for $l_{\rm mfp} = 0.2$kpc and $T=3\times 10^7 $K, and is larger than $\kappa_{\rm diff}$ in most of the parameter space. Figure \ref{fig:kappa_different_gamma} therefore shows that $\Phi < 1$ is plausible in the ICM, and so the CRBI is not completely suppressed by CR diffusion. 

\section{Conclusions}\label{sec:conclusions}
In \cite{kqs2020} we showed that streaming CRs destabilise sound waves in the low-collisionality ICM. The instability arises because the Alfv\'en speed in low-collisionality plasmas depends on the pressure anisotropy of the thermal gas (eq. \ref{eq:va_mod}). This introduces a new unstable form of coupling between CRs and the thermal plasma. In this work, we showed that Alfv\'enically streaming CRs in a gravitationally stratified medium also destabilise a pressure-balanced mode, more specifically the CR entropy mode. We term this the cosmic ray buoyancy instability (CRBI) because it is the combined action of CR streaming and gravity (buoyancy) that drives the instability. CR entropy modes are highly compressible on small scales (Figure \ref{fig:mode}), which drives them unstable in a gravitational field. In the limit of pure CR streaming (no diffusion), there likely is no threshold for the CRBI (see discussion in Sections \ref{sec:short} and \ref{sec:icm}). The fastest growth occurs at short wavelengths, where the mode is highly compressible, with growth rates of order $\eta \beta^{1/2} \omega_{\rm ff}$ (eq. \ref{eq:gamma}) where $\eta = p_c /p_g$, $\beta=8\pi p_g / B^2$ and $\omega_{\rm ff}$ is the free-fall frequency. Our results show that CR streaming in cluster plasmas is a dramatically unstable process and that CR physics is important for understanding wave propagation in the ICM, even for subdominant CR pressures.

We gave a physical overview of the CRBI in Section \ref{sec:prelims}. Instability arises due to gravity acting on the mode's density fluctuations. In standard buoyancy instabilities, such as thermal convection in stars or the magneto-thermal instability (MTI; \citealt{mti_balbus}) and the heat-flux-driven buoyancy instability (HBI; \citealt{hbi}) in clusters,  the density fluctuations are due to the background stratification of the plasma. Notably, in the CRBI the density fluctuations at short wavelengths are due to the combined action of CR streaming and pressure balance, independent of the background stratification. We complemented the qualitative physical picture from Section \ref{sec:prelims} with a quantitative dispersion-relation calculation in Section \ref{sec:grav}, and showed growth rates and mode properties for a wide range of physical parameters in Figures \ref{fig:mode}--\ref{fig:lf}. 

\subsection{Relationship to other instabilities}
Previous work on dilute cluster plasmas showed that anisotropic conduction leads to buoyancy instabilities, the MTI and HBI. Figure \ref{fig:hbi_mti} shows that these instabilities dominate growth rates at long wavelengths even in the presence of CRs, if the gas scale height is significantly larger than the thermal mean free path, as is the case in cluster cores. The CR-driven instability operates on small scales, precisely where the heat-flux-driven buoyancy instabilities are stable due to magnetic tension. The MTI/HBI and the CRBI of this paper can thus operate simultaneously in cluster plasmas. However, we note that the scale separation between the MTI/HBI and the CRBI is not always so clear: in the more dilute cluster outskirts, where the thermal mean free path is significantly larger, the CRBI can have significant growth rates (of order the free-fall frequency for plausible parameters) even for long-wavelength $kH \sim 1$ modes (Figure \ref{fig:lf}). 

\begin{figure}
  \centering
    \includegraphics[width=0.49\textwidth]{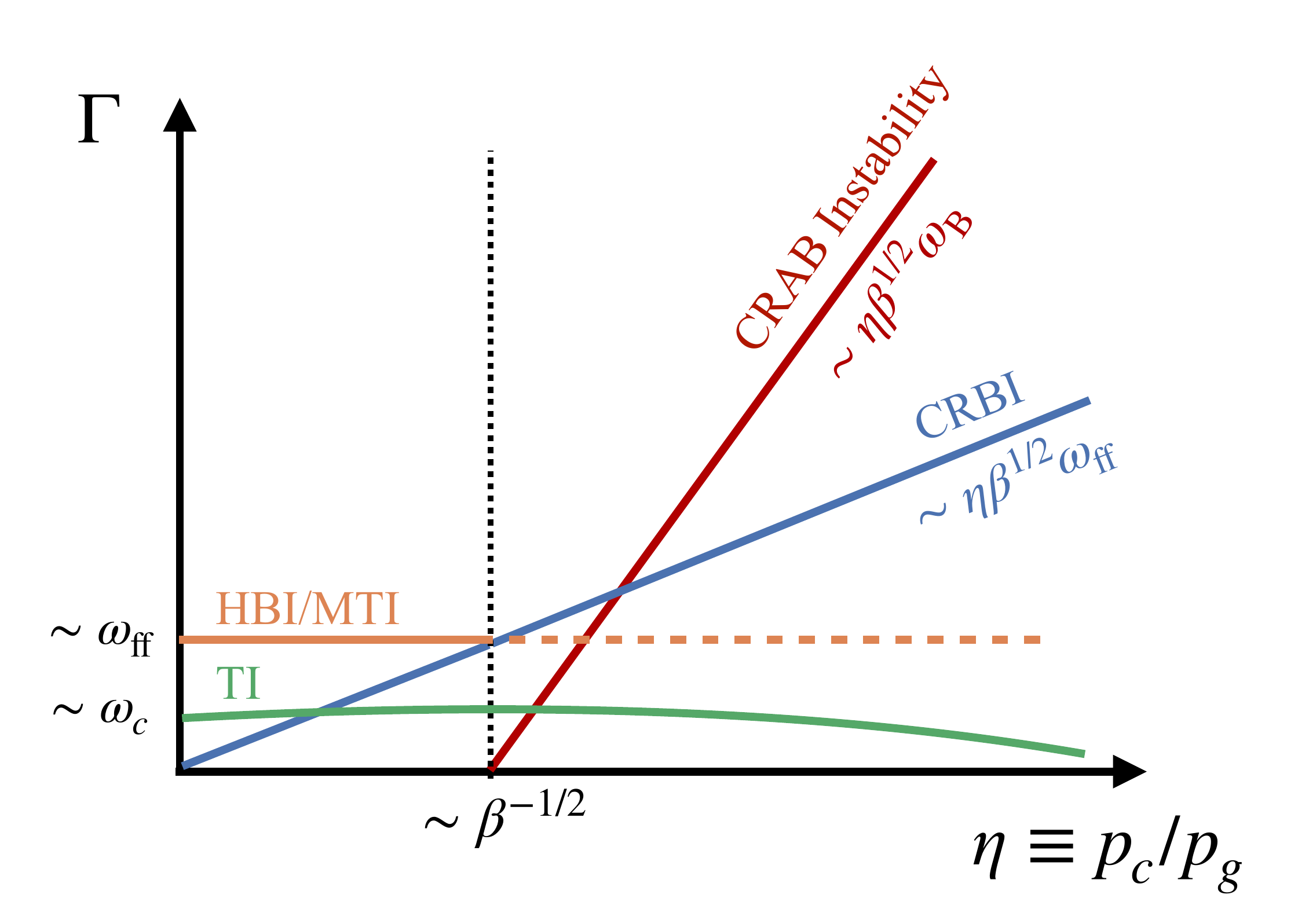}
  \caption[Schematic overview of instabilities in the ICM as a function of CR pressure.]{Overview of instabilities in dilute ICM plasmas as a function of $\eta=p_c/p_g$. For small $\eta$, the HBI/MTI have the highest growth rates. For $\eta \gtrsim \beta^{-1/2}$, the CRBI has growth rates larger than the HBI/MTI, and the CRAB instability of sound waves is excited. We include approximate growth rates of the CRBI and CRAB instability ($\Gamma \sim \eta \beta^{1/2} \omega_{\rm B}$ for the CRAB instability is valid above the instability threshold and as long as $\Gamma \ll \omega_s$; \citealt{kqs2020}). CRs do not significantly affect  thermal-instability (TI) growth rates for $\eta \lesssim 1$ (\citealt{kq2020}). The CRAB instability drives the fastest growing mode for $\eta \gtrsim \beta^{-1/2}$. However, while the CR entropy modes have smaller growth rates, they also have smaller group speeds and so remain in the region in which they are excited for longer, potentially leading to larger overall amplification. \label{fig:phase_diagram}}
\end{figure}

In Figure \ref{fig:phase_diagram}, we summarise how the CRBI and the Cosmic Ray Acoustic Braginskii (CRAB) instability from \cite{kqs2020} compare to previously identified instabilities that may operate in ICM plasmas. We sketch representative growth rates due to the different instabilities as a function of $\eta$. For small $\eta$, the HBI/MTI are the fastest growing instabilities operating in the ICM (at large CR pressures we use a dashed line for HBI/MTI because CRs may suppress the HBI, and the CRBI and HBI/MTI can be associated with the same mode; see Figure \ref{fig:hbi_mti}). For $\eta \gtrsim \beta^{-1/2}$ (recall that $\beta \gg 1$ in the ICM), the growth rate of short-wavelength CR entropy modes driven compressible by CR streaming becomes comparable to or larger than $\omega_{\rm ff}$. For $\eta \gtrsim \beta^{-1/2}$ the CRAB instability of sound waves is also excited (\citealt{kqs2020}). The impact of CRs on thermal-instability (TI) growth rates is modest (\citealt{kq2020}). The CRAB instability generally drives the fastest growing mode. This, however, does not necessarily mean that for large $\eta$ the nonlinear dynamics are dominated by the CRAB instability, as the saturation of both CR-driven instabilities remains unclear and is the subject of ongoing work. In particular, while the unstable CR entropy modes have smaller growth rates, they also have smaller group speeds and so remain in the region in which they are excited for longer. This is especially true at high $\beta$: waves propagating at the Alfv\'en speed with growth rates of order $\omega_{\rm ff}$ undergo several e-foldings over the distance of one gas scale height. 

\subsection{The CRBI and CRAB instability in cluster cores}

Heating by streaming CRs may balance cooling in the inner regions of cluster cores (\citealt{guo08}; \citealt{jp_1}; \citealt{jp_2}). For a cooling rate $\omega_c$, this requires CR pressures  of order (\citealt{kq2020}),
\begin{equation}
    \eta \sim \beta^{1/2} \frac{\omega_c}{\omega_{\rm ff}}.
\end{equation}
The CRAB instability and the CRBI become important for $\eta \beta^{1/2} \gtrsim 1$ and therefore destabilise a CR-heated medium if,
\begin{equation}
    \eta \beta^{1/2} \sim \beta \frac{\omega_c}{\omega_{\rm ff}} \gtrsim 1.
\end{equation} 
Observations suggest that $\omega_{\rm ff}/ \omega_c \gtrsim 10$ in cluster cores (e.g., \citealt{mcdonald2010}, \citealt{hogan17}). A CR-heated medium is therefore plausibly unstable to the CRAB instability and the CRBI for $\beta \gtrsim 10$, a condition that is likely satisfied in the ICM.  

We also note that the CRAB and CR buoyancy instabilities may have, to some extent, similar observational appearances. In particular, although CR entropy modes are pressure-balanced, they are compressible and involve finite gas-pressure fluctuations (balanced by CR-pressure fluctuations). CR entropy modes may therefore masquerade as sound waves if only the thermal-gas fluctuations are measured. Moreover, due to their compressible nature both the CRAB instability and the CRBI may evolve into shock-like structures that resemble the weak shocks observed in the Perseus cluster (\citealt{fabian03}; \citealt{fabian06}).

In the standard picture, AGN in cool cluster cores excite sound waves and internal gravity waves via the time dependence of the AGN jet and the buoyant motion of radio bubbles into the ICM. \cite{kqs2020} and this work suggest that  waves can also be excited by the CR pressure gradient that the bubbles provide. Future simulations will address the nonlinear evolution of the CR-driven instabilities. There are two important stages that are crucial for the nonlinear evolution and saturation: when the amplitudes become large enough to locally flatten the CR pressure gradient and shut off CR streaming ($\delta p_c / p_c \sim 1/kH$; though this does not necessarily shut off the instability, see \citealt{navin_2021_staircase}) and  when the amplitudes become large enough for the pressure anisotropy to excite kinetic microinstabilities such as the mirror (\citealt{b66}; \citealt{h69}) and firehose (\citealt{r56}; \citealt{c58}; \citealt{p58}) instabilities (which occur when $|\Delta p| \sim B^2 / 4 \pi$). Upcoming work will address how this additional physics, which is not part of the linear analysis presented here, affects the evolution of the CR-driven instabilities and their impact on the ICM. 

\subsection{Dependence on CR transport physics}

Both the CRBI considered in this work and the CRAB instability in \cite{kqs2020} are driven by CR streaming at the Alfv\'en speed. The instabilities therefore operate only if the bulk of CRs in the ICM are self-confined, rather than scattered by an extrinsic turbulent cascade of magnetic fluctuations. According to current theoretical models of MHD turbulence, CR scattering by Alfv\'enic turbulence is likely negligible, due to the anisotropy of the cascade (\citealt{chandran_scattering}). The MHD weak cascade of fast modes may be isotropic and more efficient at scattering CRs, and has been proposed as an alternative to self-confinement (\citealt{yan_lazarian_2004}). However, because the weak cascade of fast modes is strongly damped in dilute high-$\beta$ plasmas, and may generally be suppressed by wave steepening (\citealt{kadomtsev1973acoustic}; \citealt{kq2022}), scattering of energetically important GeV CRs by fast modes is likely suppressed in the high-$\beta$ ICM. Self-confinement and streaming transport are therefore plausible. 

The finite CR mean free path in the frame moving with the self-excited Alfv\'en waves necessarily implies a correction to the pure Alfv\'enic streaming model. Significant CR diffusion resulting from this correction can suppress the CRBI (Figure \ref{fig:diff}). However, the magnitude and nature of the correction to Alfv\'enic streaming remains uncertain. In particular, the form of the transport correction turns out to be rather peculiar, as it corresponds to neither streaming nor diffusion (\citealt{skilling71}; \citealt{wiener2013}; \citealt{kq2022}). We attempted to quantify the magnitude of the \textit{diffusive} part of the transport correction, i.e. the contribution that may suppress the CRBI, in Section \ref{sec:diffusion}, which was based on the calculation from Appendix \ref{sec:diff_appendix}.  Figure \ref{fig:kappa_different_gamma} shows that the CRBI is usually not suppressed by CR diffusion for expected ICM conditions. It would also be valuable to carry out a more complete calculation -- based on CR kinetic theory -- to test the conclusions of our simplified fluid treatment (although it is worth noting that existing theories of CR transport are quite uncertain and have difficulties explaining CR measurements in the Milky Way; e.g., \citealt{kq2022}, \citealt{hopkins_sc_et_problems}).

\section*{Data Availability}
The calculations from this article will be shared on reasonable request to the corresponding author.

\section*{Acknowledgements}
We thank Matthew Kunz for useful comments. This research was supported in part by the National Science Foundation under Grant No. NSF PHY-1748958, by NSF grant AST-2107872 and a Simons Investigator award from the Simons Foundation. Support for J.S. was provided by Rutherford Discovery Fellowship RDF-U001804, which is managed through the Royal Society Te Ap\={a}rangi.

\bibliographystyle{mnras}

\bibliography{crbi}


\appendix

\section{Landau-fluid closure for low-collisionality plasmas} \label{app:lf}
Here we provide the kinetic MHD equations and the Landau-fluid closure for the heat fluxes used in Section \ref{sec:collisionless} and Figure \ref{fig:lf}. The kinetic MHD evolution equations for the pressures perpendicular and parallel to the magnetic field are (\citealt{cgl}),
\begin{equation}\label{eq:p_perp}
\frac{\partial p_\perp}{\partial t} + \bm{\nabla \cdot} (p_\perp \bvrm) + p_\perp \bm{\nabla \cdot} \bvrm + \bm{\nabla \cdot }(q_\perp \bm{\hat{b}}) + q_\perp \bm{\nabla \cdot \hat{b}} = p_\perp \mathbf{ \hat{b}\hat{b} : \nabla \bvrm }  - \frac{1}{3} \nu_{\rm ii} \Delta p, 
\end{equation}
\begin{gather}
\begin{aligned}\label{eq:p_par}
\frac{\partial p_\parallel}{\partial t} + \bm{\nabla \cdot} (p_\parallel \bvrm) + \bm{\nabla \cdot }(q_\parallel \bm{\hat{b}}) - 2 q_\perp \bm{\nabla \cdot \hat{b}} = & -2 p_\parallel \mathbf{ \hat{b}\hat{b} : \nabla \bvrm}  + \frac{2}{3} \nu_{\rm ii} \Delta p \\ &- 3(\gamma - 1) \bm{{\rm v_{st}} \cdot \nabla }p_c, 
\end{aligned}
\end{gather}
where we made the somewhat uncertain assumption that CR heating is predominantly in the direction parallel to the magnetic field. This is motivated by the fact that CR heating is due to the excitation of parallel-propagating modes, although we note that this is not true if damping by Alfv\'enic turbulence dominates, which acts to shear the waves to high $k_\perp$. This choice does not, however, significantly affect the results. The above equations are not yet complete, as the heat fluxes are still undetermined. In the Landau-fluid closure, the heat fluxes are set such that the linear behaviour of the fluid model approximately matches the linear response of the fully kinetic thermal plasma (\citealt{snyder97}). The Landau-fluid closure has been popular for modeling collisionless plasmas, as it  recovers the fully kinetic linear damping rates (e.g. linear Landau damping of ion acoustic waves) and instabilities (e.g. MRI) of all MHD modes. A convenient form for the heat fluxes, which recovers Braginskii MHD in the collisional limit, is given by (\citealt{snyder97}),
\begin{equation} \label{eq:qperp}
    q_\perp = - \frac{2 c_{s \parallel}^2}{\sqrt{2\pi}|k_\parallel| c_{s\parallel} + \nu_{\rm ii}} \Big[\rho \nabla_\parallel \Big(\frac{p_\perp}{\rho}\Big) - p_\perp \Big(1- \frac{p_\perp}{p_\parallel} \Big) \frac{\nabla_\parallel B}{B} \Big],
\end{equation}
\begin{equation} \label{eq:qpar}
    q_\parallel = - \frac{8 c_{s \parallel}^2}{\sqrt{8\pi}|k_\parallel| c_{s\parallel} + (3 \pi -8)\nu_{\rm ii}} \rho \nabla_\parallel \Big(\frac{p_\parallel}{\rho}\Big),
\end{equation}
where $c_{s\parallel} = \sqrt{p_\parallel / \rho}$.
In Section \ref{sec:collisionless} and Figure \ref{fig:lf} we use the linearised versions of \eqref{eq:p_perp}--\eqref{eq:qpar} instead of the linearised Braginskii MHD equations \eqref{eq:dpg} and \eqref{eq:dDp_Brag}.

\section{CR Diffusion Coefficient in Self-Confinement Theory} \label{sec:diff_appendix}

\begin{figure}
  \centering
    \includegraphics[width=0.45\textwidth]{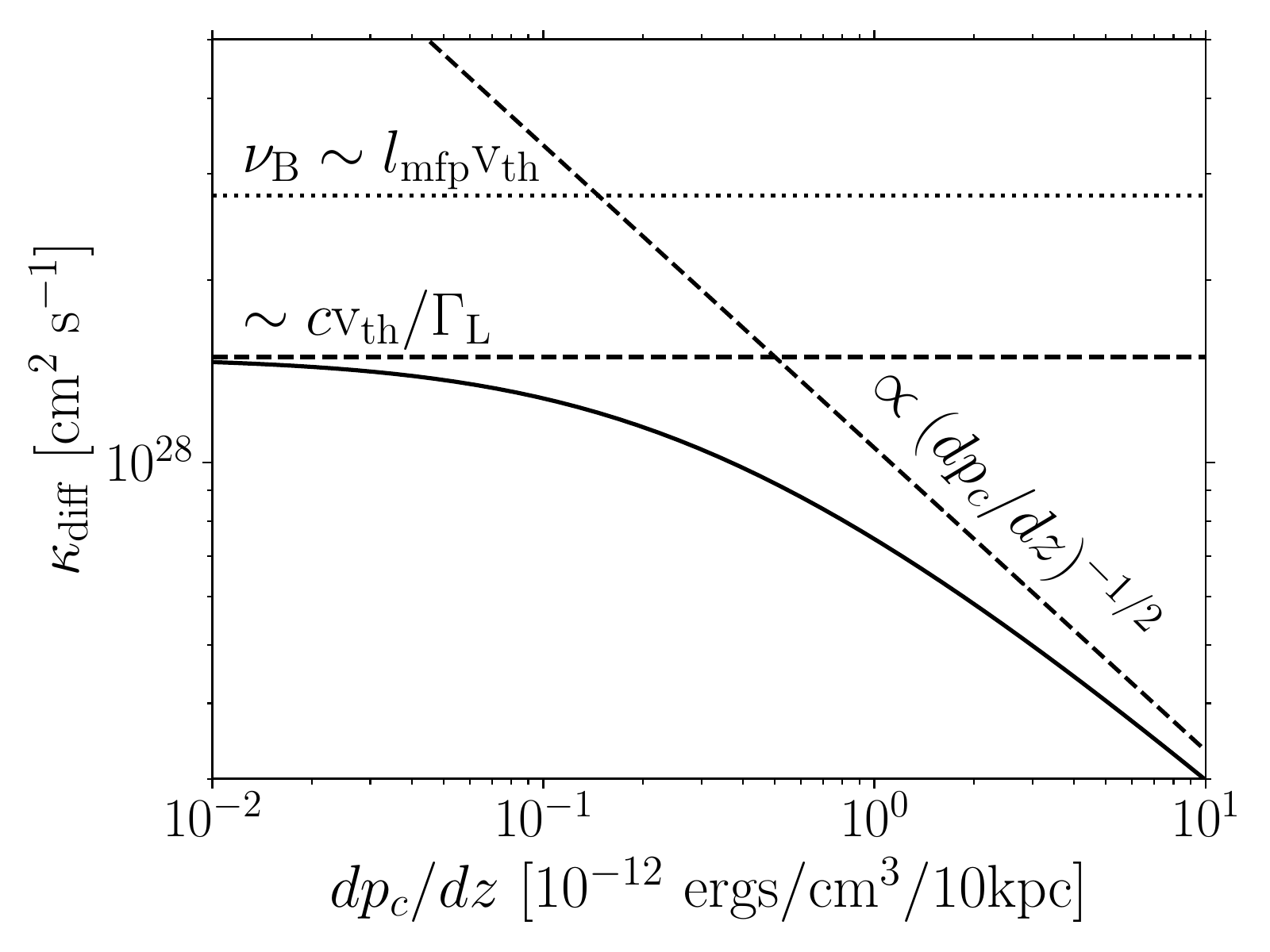}
  \caption[Diffusive component of the CR diffusion coefficient in self-confinement theory as a function of the CR pressure gradient.]{ $\kappa_{\rm diff}$ as a function of the CR pressure gradient for a linear damping rate $\Gamma_{\rm L} = 10^{-10} \ s^{-1}$, where  $\kappa_{\rm diff}$ is the component of the diffusion coefficient $\kappa$ that gives rise to actually diffusive behaviour. We estimate $\kappa_{\rm diff}$ from self-confinement theory by computing the total diffusion coefficient $\kappa$ from eq. \eqref{eq:quad_balance} and  subtracting $\kappa_{\rm st}$ (eq. \ref{eq:diff_coeff_st}) in order to not include the non-diffusive correction when linear damping dominates. While this method of computing $\kappa_{\rm diff}$ is not exact, it correctly recovers the diffusive correction in the two asymptotic limits shown by the dashed lines (equations \ref{eq:diff_coeff_D} and \ref{eq:diff_coeff}). This suggests that the solid line is a reasonable approximation of the diffusive correction to Alfv\'enic streaming.  \label{fig:kappa}}
\end{figure}

In this section we provide a heuristic calculation of the CR diffusion coefficient in self-confinement theory (a similar calculation can be found in \citealt{hopkins_testing_cr}). One challenge in this calculation is that leading-order corrections to Alfv\'enic streaming are often not diffusive. Instead they are better described by a (super-Alfv\'enic) streaming or sink term  (this is the case when linear damping of Alfv\'en waves dominates; \citealt{skilling71}; \citealt{wiener2013}; \citealt{kq2022}). For our linear analysis calculation, we are mainly interested in the leading-order \emph{diffusive} correction, which is more likely to suppress the instability than a streaming/sink term.

We calculate the amplitude of waves excited by the CR streaming instability, and the resulting CR scattering frequency, by equating Alfv\'en-wave growth and damping. We consider a steady state with
\begin{equation} \label{eq:balance}
    (\Gamma_{\rm L} + \Gamma_{\rm NL}) \frac{\delta B^2}{4 \pi} = | \bm{ {\rm  v_A} \cdot \nabla} p_c |,
\end{equation}
where we split the wave damping into a linear and nonlinear part ($\propto \delta B^2$). $\Gamma_{\rm L}$ is the sum of all linear damping contributions, turbulent (\citealt{farmer_goldreich}), linear-Landau (\citealt{wiener_hib}), ion-neutral and dust (\citealt{squire_dust_2021}) damping, although the latter two are likely not important in the hot and dilute ICM. $\Gamma_{\rm NL}$ is the nonlinear Landau damping rate (\citealt{lee_volk_1973}; \citealt{kulsrud_book}), $\Gamma_{\rm NL} = \gamma_{\rm NL} (\delta B / B)^2$, where $\gamma_{\rm NL} \sim k {\rm v_{\rm th}}$, $k$ is the wavenumber of Alfv\'en waves resonant with $\sim$ GeV CRs and ${\rm v_{th}}$ is the ion thermal speed. Equation \ref{eq:balance} becomes a quadratic equation for the wave amplitude    
\begin{equation}\label{eq:quad_balance}
    \Big( \frac{\delta B^2}{B^2} \Big)^2 + \frac{\Gamma_{\rm L}}{\gamma_{\rm NL}} \frac{\delta B^2}{B^2} - \frac{\Gamma_{\rm A}}{2 \gamma_{\rm NL}} \frac{p_c}{\epsilon_{\rm B}} = 0,
\end{equation}
where $\Gamma_{\rm A} \equiv | \bm{ {\rm  v_A} \cdot \nabla} p_c | / p_c$ is the inverse of the Alfv\'en crossing time and $\epsilon_{\rm B}$ is the magnetic-field energy density. We first consider the the limit $\Gamma_{\rm L} \gg \Gamma_{\rm NL}$. \eqref{eq:quad_balance} can then be solved perturbatively to yield
\begin{equation}
    \frac{\delta B^2}{B^2} \approx \frac{\Gamma_{\rm A} p_c}{2 \Gamma_{\rm L} \epsilon_{\rm B}} \Big(1 - \frac{\gamma_{\rm NL}}{ \Gamma_{\rm L}} \frac{\Gamma_{\rm A} p_c}{2 \Gamma_{\rm L} \epsilon_{\rm B}} \Big).
\end{equation}
The pitch-angle scattering rate of GeV CRs is $\nu_{\rm CR} \sim \Omega_0 \delta B^2 / B^2$, which corresponds to a diffusion coefficient
\begin{equation} \label{eq:diff_coeff0}
    \kappa \sim \frac{c^2}{\nu_{\rm CR}} \approx \kappa_{\rm st} + \kappa_{\rm diff},
\end{equation}
where,
 \begin{equation}\label{eq:diff_coeff_st}
     \kappa_{\rm st} =  \frac{c^2}{\Omega_0} \frac{2 \Gamma_{\rm L} \epsilon_{\rm B}}{\Gamma_{\rm A} p_c},
 \end{equation}
 and
  \begin{equation}\label{eq:diff_coeff_D}
     \kappa_{\rm diff} =  \frac{c^2}{\Omega_0} \frac{\gamma_{\rm NL}}{ \Gamma_{\rm L}} \sim \frac{c {\rm v}_{\rm th}}{\Gamma_{\rm L}}.
 \end{equation}
$\kappa_{\rm st} \propto | \bm{ {\rm  v_A} \cdot \nabla} p_c |^{-1}$ reflects the well-known result that for purely linear damping rates the diffusion coefficient is inversely proportional to the CR pressure gradient, so that the diffusion term $\bm{\nabla \cdot}(\kappa \bm{\hat{b} \hat{b} \cdot \nabla}p_c)$ ends up not being diffusive at all and is better described as a (super-Alfv\'enic) streaming or sink term (\citealt{skilling71}; \citealt{wiener2013}; \citealt{kq2022}). By contrast, $\kappa_{\rm diff}$ is independent of the CR pressure gradient and is therefore a regular diffusion coefficient. 

Conversely, if nonlinear Landau damping dominates, the CR diffusion coefficient is to leading order (from eq. \ref{eq:quad_balance}):
\begin{equation} \label{eq:diff_coeff}
    \kappa_{\rm NL} \approx \kappa_{\rm diff}   \approx  \frac{c^2}{\Omega_0} \Big( \frac{2 \gamma_{\rm NL} \epsilon_{\rm B}}{\Gamma_{\rm A} p_c} \Big)^{1/2}.
\end{equation}
$\kappa_{\rm NL} \propto | \bm{ {\rm  v_A} \cdot \nabla} p_c |^{-1/2}$ and so we end up with a term that is again not diffusive in the usual sense. However, in linear theory with a background CR pressure gradient, $\bm{\nabla \cdot}(\kappa \bm{\hat{b} \hat{b} \cdot \nabla}p_c)$ still gives a term $\propto \kappa k^2$ (where $\kappa$ depends on the background gradient) and is therefore linearly diffusive. 

In the high-$\beta$ ICM, linear Landau damping (\citealt{wiener_hib}) is likely the most important linear damping rate. For turbulence injected on $\sim 10$ kpc (common scale of the radio bubbles) with perturbations comparable to the Alfv\'en speed, the damping rate of $k \sim r_L^{-1}$ Alfv\'en waves excited by GeV CRs (where $r_L$ is the GeV CR gyroradius) is
\begin{equation}
    \Gamma_{\rm L} \sim \frac{0.4 \vrm_{\rm th}}{(r_L L_{\rm turb})^{1/2}} \sim 10^{-10} \ {\rm s^{-1}} \frac{\vrm_{\rm th}}{10^8 {\rm cm \ s^{-1}}} \Big(\frac{L_{\rm turb}}{10 \ {\rm kpc}}\Big)^{-1/2} \Big(\frac{B}{1 \ {\rm \mu G}} \Big)^{1/2}.
\end{equation}

We plot $\kappa_{\rm diff}$, i.e. the component of the diffusion coefficient $\kappa$ that gives rise to actually diffusive behaviour, as a function of the CR pressure gradient in Figure \ref{fig:kappa} (see also Figure \ref{fig:kappa_different_gamma} for a different version of this plot). We calculate $\kappa_{\rm diff}$ by computing the total diffusion coefficient $\kappa$ from eq. \eqref{eq:quad_balance} and  subtracting $\kappa_{\rm st}$ (eq. \ref{eq:diff_coeff_st}) in order to not include the non-diffusive correction when linear damping dominates. Simply subtracting $\kappa_{\rm st}$ to obtain the diffusive correction is not exact. However, it correctly recovers the two asymptotic limits (equations \ref{eq:diff_coeff_D} and \ref{eq:diff_coeff} and the dashed lines in Figure \ref{fig:kappa}). This suggests that the solid line in Figure \ref{fig:kappa} is a reasonable approximation of the diffusive correction to Alfv\'enic streaming.  

For large CR pressure gradients, the streaming instability reaches large amplitudes (for a fixed linear damping rate) and nonlinear Landau damping is more important  than linear damping mechanisms. The resulting diffusion coefficients are $\propto (dp_c/dz)^{-1/2}$. For small CR pressure gradients, linear damping dominates as the amplitudes reached by the streaming instability are not large enough for nonlinear Landau damping to be important. The diffusion coefficient is constant and approximately given by eq. \eqref{eq:diff_coeff_D}. The horizontal dotted line in Figure \ref{fig:kappa} is the anisotropic viscosity of the thermal gas for $l_{\rm mfp} \sim 0.2 \ {\rm kpc}$ and $T = 3 \times 10^7$ K. It is therefore plausible to expect $\Phi < 1$ in cluster cores and the CRBI remains active, though is likely partially suppressed (Figure \ref{fig:diff}). 

\bsp	
\label{lastpage}
\end{document}